\documentclass[12pt]{emulateapj}

\usepackage{amsmath,amssymb}  
\usepackage{graphicx}

\usepackage{color}

\newcommand{\cgm}{$ \xi_{\rm gm} $}

\newcommand{\ncen}{\langle N_{\rm cen} \rangle}
\newcommand{\nsat}{\langle N_{\rm sat} \rangle}
\newcommand{\ntot}{\langle N_{\rm tot} \rangle}
\newcommand{\mmin}{M_{\rm min}}

\newcommand{\asat}{\alpha_{\rm sat}}

\def\fshmr{f^{-1}_{\textsc{shmr}}(M_\ast^{t_1})}

\def\ntotb{\langle N_{\rm tot}(M_h|M_{*}^{t_1},M_{*}^{t2})\rangle}

\slugcomment{ }
 
\shorttitle{Combining dark matter probes}
\shortauthors{A.\ Leauthaud et al.}

\begin{document}
  

 \title{A theoretical framework for combining techniques that probe
   the link between galaxies and dark matter.}




\author{Alexie Leauthaud\altaffilmark{1,2},
Jeremy Tinker\altaffilmark{3},
Peter S. Behroozi\altaffilmark{4},
Michael T. Busha\altaffilmark{4,5},
Risa H. Wechsler\altaffilmark{4}}

\email{asleauthaud@lbl.gov}

\altaffiltext{1}{Lawrence Berkeley National Lab, 1 Cyclotron Road,
  Berkeley, CA 94720, USA}

\altaffiltext{2}{Berkeley Center for Cosmological Physics, University
  of California, Berkeley, CA 94720, USA}

\altaffiltext{3}{Center for Cosmology and Particle Physics, Department
  of Physics, New York University, 4 Washington Place, New York, NY
  10003}

\altaffiltext{4}{Kavli Institute for Particle Astrophysics and
  Cosmology; Physics Department, Stanford University, and SLAC
  National Accelerator Laboratory, Stanford CA 94305}

\altaffiltext{5}{Institute for Theoretical Physics, Department of
  Physics, University of Zurich, CH-8057, Switzerland}

  
\begin{abstract}We develop a theoretical framework that combines
  measurements of galaxy-galaxy lensing, galaxy clustering, and the
  galaxy stellar mass function in a self-consistent manner. While
  considerable effort has been invested in exploring each of these
  probes individually, attempts to combine them are still in their
  infancy.  These combinations have potential to elucidate the
  galaxy-dark matter connection and the galaxy formation physics that
  is responsible for it, as well as to constrain cosmological
  parameters, and to test the nature of gravity. In this paper, we
  focus on a theoretical model that describes the galaxy-dark matter
  connection based on standard halo occupation distribution
  techniques.  Several key modifications enable us to extract
  additional parameters that determine the stellar-to-halo mass
  relation and to simultaneously fit data from multiple probes while
  allowing for independent binning schemes for each probe. We
  construct mock catalogs from numerical simulations to investigate
  the effects of sample variance and covariance for each
  probe. Finally, we analyze how trends in each of the three
  observables impact the derived parameters of the model. In
  particular, we investigate various features of the observed galaxy
  stellar mass function(low-mass slope, ``plateau'', knee, and
  high-mass cut-off) and show how each feature is related to the
  underlying relationship between stellar and halo mass. We
  demonstrate that the observed ``plateau'' feature in the stellar mass
  function at $M_*\sim 2\times10^{10}$ M$_{\odot}$ is due to the
  transition that occurs in the stellar-to-halo mass relation at
  $M_h\sim 10^{12}$ M$_{\odot}$ from a low-mass power-law regime to a
  sub-exponential function at higher stellar mass.\end{abstract}
 

 
 


\section{Introduction}

Improved measurements of the link between galaxies and the dark matter
distribution will benefit a variety of cosmological applications but
will also provide important clues about the role that dark matter
plays in galaxy formation process. Although multiple techniques have
been developed for this purpose, no single method has yet emerged as
the ultimate tool and all suffer from various drawbacks. The goal of
this paper is to develop the theoretical foundations required to
combine multiple probes into a single tool that will provide more
powerful constraints on the galaxy-dark matter connection. This paper
extends and complements a growing body of work on this topic
\citep[][]{Seljak:2000,Guzik:2001,Guzik:2002,Berlind:2002,Tasitsiomi:2004,Mandelbaum:2005a,Mandelbaum:2006c,Yoo:2006,Cacciato:2009,Tinker:2011}.


At present, there are only two observational techniques capable of
{\em directly} probing the dark matter halos of galaxies out to large
radii (above 50 kpc): galaxy-galaxy lensing \citep[e.g.,][]{
  Brainerd:1996,McKay:2001, Hoekstra:2004, Sheldon:2004,
  Mandelbaum:2006, Mandelbaum:2006c,
  Heymans:2006,Johnston:2007,Leauthaud:2010} and the kinematics of
satellite galaxies \citep[][]{McKay:2002, Prada:2003, Brainerd:2003,
  van-den-Bosch:2004, Conroy:2007, Becker:2007, Norberg:2008,
  More:2009, More:2011}.  Galaxy-galaxy lensing (hereafter ``g-g
lensing'') utilizes subtle distortions induced in the shapes and
orientations of distant background galaxies in order to measure
foreground mass distributions. The satellite kinematic method uses
satellite galaxies as test particles to trace out the dark matter
velocity field. Neither method can probe the halos of individual
galaxies. Instead, both techniques must stack an ensemble of
foreground galaxies in order to extract a signal. Nonetheless, with
the advent of data-sets large enough to provide statistically
significant samples, improvements in photometric redshift techniques,
and spectroscopic follow-up programs, both methods have emerged as
powerful probes of the galaxy-dark matter connection and have truly
evolved into mature techniques over the last decade.


In addition to these two direct probes, there are also several popular
indirect methods to infer the galaxy-dark matter connection from the
statistics of galaxy clustering. For example, numerous authors have
employed a statistical model to describe the probability distribution
$P(N|M_h)$ that a halo of mass $M_h$ is host to N galaxies above some
threshold in luminosity or stellar-mass. This statistical model,
commonly known as the halo occupation distribution (HOD), has been
considerably successful at interpreting the clustering properties of
galaxies \citep[e.g.,][]{Seljak:2000, Peacock:2000, Scoccimarro:2001,
  Berlind:2002, Bullock:2002, Zehavi:2002, Zehavi:2005, Zheng:2005,
  Zheng:2007, Tinker:2007,Wake:2011,Zehavi:2010,White:2011}. The HOD
provides a description of the spatial distribution of galaxies at all
scales, but it is usually inferred observationally by modeling
measurements of the two-point correlation function of galaxies,
$\xi_{gg}(r)$. Since they were introduced a decade ago, HOD models
have progressively increased in fidelity and complexity owing to
stronger observational constraints but also to the availability of
larger, high-resolution cosmological N-body simulations of the dark
matter. For example, analytical descriptions of the form and evolution
of the halo mass function and the large-scale halo bias, both of which
are key ingredients for HOD models, are approaching percent-level
precision \citep[e.g.,][]{Tinker:2008,Tinker:2010}. A variety of
extensions to the basic HOD framework have also been proposed. For
example, the conditional luminosity function $\Phi (L|M_h)dL$
specifies the average number of galaxies of luminosity $L\pm dL/2$
that reside in a halo of mass $M_h $ \citep[e.g.,][]{Yang:2003,
  van-den-Bosch:2003, Vale:2004, Cooray:2006, van-den-Bosch:2007,
  Vale:2008} and the conditional stellar mass function
$\Phi(M_{*}|M_h)dM_*$ describes the average number of galaxies with
stellar masses in the range $M_{*}\pm dM_{*}$ as a function of host
halo mass $M_h$
\citep[e.g.,][]{Yang:2009,Moster:2010,Behroozi:2010}. Furthermore, a
number of studies are also starting to take into account, not only the
simple expectation values of the underlying relations, but also the
scatter between the observable and halo-mass
\citep[e.g.,][]{More:2011,Behroozi:2010,Moster:2010}, a crucial
ingredient for a complete description of the galaxy-dark matter
connection.

Finally, halo mass constraints from the galaxy stellar mass function
(hereafter ``SMF'') have also been derived by assuming that there is
there is a monotonic correspondence between halo mass (or circular
velocity) and galaxy stellar mass (or luminosity)
\citep[e.g.,][]{Kravtsov:2004,Vale:2004,Tasitsiomi:2004,Vale:2006a,Conroy:2009,Drory:2009,Moster:2010,Behroozi:2010,Guo:2010}. This
particular technique, often referred to as ``abundance matching'', is
economic in terms of data requirements since it only considers the
observed stellar mass (or luminosity) function. However, prior
knowledge about the mass distribution of halos (and substructure
within those halos) from cosmological N-body simulations is necessary
as well as the assumption that field halos and subhalos of the same
halo mass contain galaxies of the same stellar mass.

While considerable effort has been invested in exploring each of these
probes individually, attempts to {\em combine} them in a fully
consistent way are still in their infancy. Nevertheless, savvy
combinations hold great potential to not only elucidate the evolution
of the galaxy-dark matter connection, and consequently the galaxy
formation physics responsible for it, but to also constrain
fundamental physics, including the cosmological model and the nature
of gravity. For example, measurements of small-scale galaxy clustering
alone do not yield cosmological constraints unless coupled with probes
that are sensitive to the mass scales of dark matter halos (e.g.,
cluster mass-to-light ratios, satellite kinematics, g-g lensing, etc.)
\citep[][]{van-den-Bosch:2003a,Tinker:2005,Seljak:2005}. In
particular, \citet[][]{Yoo:2006} and \citet[][]{Cacciato:2009} have
shown that the combination of g-g lensing and galaxy clustering is
sensitive to $\Omega_m$ and $\sigma_8$. Conceptually, this sensitivity
arises from the fact that this particular combination simultaneously
probes the shape and amplitude of the halo mass function at small
scales and the overall matter density and the bias of the galaxy
sample at large scales.

Other combinations can be sensitive to parameters in modified gravity
theories.  A generic metric theory of gravity has two scalar
potentials, $\phi$, which affects the clustering and dynamics of
galaxies, and $\psi$, which affects the lensing of light around
galaxies.  Combining probes of g-g lensing with clustering and/or
satellite dynamics allows a test of the general relativity (GR)
prediction that $\psi = \phi$ as well as the Poisson equations which
relate these potentials to the underlying density distribution. For
example, \citet[][]{Reyes:2010} have used a combination of g-g
lensing, galaxy clustering, and redshift space distortions to place
limits on possible modifications to GR on $\sim$10 Mpc scales.

Tests of gravity on smaller scales, though complicated by the fact
that structures have undergone non-linear evolution, are interesting
in several respects. First, \emph{all of our direct probes} of the
dark matter are either intrinsically limited to small scales (e.g., a
few Mpc for satellite kinematics) or have significantly larger signals
on small scales (e.g., below 10 Mpc for both cosmic shear and g-g
lensing).  Second, there are many alternative gravity theories which
predict unique and interesting modifications on these scales
\citep[][]{Smith:2009,Hui:2009,Schmidt:2010,Jain:2010}.

In this paper we develop the theoretical framework necessary to
constrain the galaxy-dark matter connection by combining measurements
of galaxy clustering, g-g lensing, and the galaxy SMF. The formalism
outlined in this paper could also be applied to model satellite
kinematics (see \citealt{More:2011}). For this work, we adopt the
standard HOD framework but with several key modifications. For
instance, a procedure often adopted in clustering studies is to fit a
set of HOD parameters (typically three to five) independently to the
clustering signal and number density for each galaxy sample. However,
adopting this strategy would require selecting a common binning scheme
for all probes. In practice, we would like to avoid using a single
binning scheme because various probes have different signal-to-noise
(S/N) requirements. We therefore modify the standard HOD model so that
we can simultaneously fit data from multiple probes while allowing for
independent binning schemes for each probe. Also, since we are
interested in the galaxy-dark matter connection, we modify the HOD
model so as to specifically include a parameterization for the
stellar-to-halo mass relation (hereafter ``SHMR''). In Leauthaud et
al. 2011b (hereafter Paper II), we demonstrate that this model
provides an excellent fit to g-g lensing, galaxy clustering, and
stellar mass function measurements in the COSMOS survey from $z=0.2$
to $z=1.0$.

For 2 deg$^2$ surveys such as COSMOS, the finite sample size of the
observational data set is also an important concern. We will present
an estimate of the sample variance using mock surveys from numerical
simulations. We will also estimate the covariance of the data for each
observational measure. We will demonstrate that this is especially
important for modeling the SMF, an effect that is usually not
incorporated into most analyses. The finite sample size in COSMOS
biases clustering measurements through the integral constraint, an
effect we will also model through our mock surveys.

The layout of this paper is as follows. To begin with, we introduce
the parametric form used to model the SHMR in
$\S$\ref{the_shmr}. Next, in $\S$\ref{hodtheory}, we present the
general HOD framework and our extensions to this model. In
$\S$\ref{application_of_model}, we show how this model can be used to
simultaneously fit g-g lensing, galaxy clustering, and SMF
measurements. In $\S$\ref{parameter_influance}, we describe the
influence of each model parameter on the three observables. We then
construct a set of mocks catalogs designed to mimic the COSMOS survey
and describe the behaviour of the covariance matrices for the three
probes in $\S$\ref{mocks}. Finally, we draw up our conclusions in
$\S$\ref{conclusions}.

We assume a WMAP5 $\Lambda$CDM cosmology with $\Omega_{\rm m}=0.258$,
$\Omega_\Lambda=0.742$, $\Omega_{\rm b}h^2=0.02273$, $n_{\rm
  s}=0.963$, $\sigma_{8}=0.796$, $H_0=72$ km~s$^{-1}$~Mpc$^{-1}$
\citep[][]{Hinshaw:2009}. Unless stated otherwise, all distances are
expressed in physical Mpc. The letter $M_h$ denotes halo mass. The
halo radius is noted $R_h$. In this paper, halo mass is defined as
$M_{200b}\equiv M(<R_{200b})=200\bar{\rho} \frac{4}{3}\pi R_{200b}^3$
where $R_{200b}$ is the radius at which the mean interior density is
equal to 200 times the mean matter density ($\bar{\rho}$). We note
however that our theoretical framework is valid for any reasonable
choice of halo definition. Stellar mass is noted $M_{*}$.


\section{The stellar-to-halo mass relation for central
  galaxies}\label{the_shmr}

To begin with, we present the mathematical function that we use to
model the SHMR and we describe the influence of each of the five
parameters that regulate its shape. We will assume that the SHMR is
specifically valid for ``central'' galaxies which are located by
definition at the center of their parent halos. Dark matter halos also
contain smaller bound density peaks that orbit around the center of
the potential well. These substructures are commonly referred to as
sub-halos; these sub-halos are the likely sites of ``satellite''
galaxies that have been accreted onto their parent halos. The
abundance matching technique commonly assumes that satellite galaxies
follow the same SHMR as centrals provided that halo mass is defined at
the epoch when satellites were accreted onto their parent halos
(M$_{\rm acc}$), rather than the current sub-halo mass
\citep[][]{Conroy:2006,Moster:2010,Behroozi:2010}. However, this
pre-supposes that satellite stellar growth occurs at a similar rate as
centrals of equivalent halo mass (that is to say with a halo mass
equal to M$_{\rm acc}$). Since one might expect that satellites and
centrals experience distinct stellar growth rates, we model central
and satellite galaxies separately in order to keep our model as
general as possible.

In $\S$ \ref{ncen_model}, we will show how the SHMR can be used to
predict the central occupation function and then we will introduce the
model for satellite galaxies in $\S$ \ref{model_nsat}.

\subsection{Functional form for the SHMR}

Let us consider the conditional stellar mass function (the analog of
the conditional luminosity function) which represents the number of
galaxies with $M_*$ in the range $M_*\pm dM_*/2$ at fixed halo mass
and is noted $\Phi(M_{*}|M_h)$
\citep[e.g.,][]{Yang:2009,Moster:2010,Behroozi:2010}. The conditional
stellar mass function can be divided into a central component and a
satellite component:
$\Phi(M_{*}|M_h)=\Phi_c(M_{*}|M_h)+\Phi_s(M_{*}|M_h)$. $\Phi_c(M_{*}|M_h)$
is the conditional stellar mass function for central galaxies, and it
will be our mathematical representation of the SHMR. Note that in our
model, the halo mass in the term $\Phi_s(M_{*}|M_h)$ refers to the
host halo mass.

In addition to the shape and evolution of the mean SHMR, astrophysical
processes are expected to induce an intrinsic scatter in stellar mass
at fixed halo mass which it is important to take into consideration
when defining a functional form for $\Phi_c(M_{*}|M_h)$. Another non
negligible source of scatter can be the measurement error associated
with the determination of stellar masses. In the absence of strong
observational or theoretical guidance for the form and magnitude of
the total scatter (intrinsic plus measurement), we adopt a stochastic
model where $\Phi_c(M_{*}|M_h)$ is a log-normal probability
distribution function (hereafter ``PDF'') with a log-normal
scatter\footnotemark[1]\footnotetext[1]{Scatter is quoted as the
  standard deviation of the logarithm base 10 of the stellar mass at
  fixed halo mass.} noted $\sigma_{\rm log M_{*}}$. Since we have
assumed a log-normal functional form, $\Phi_c(M_{*}|M_h)$ can be
written as:

\begin{displaymath}
\Phi_c(M_{*}|M_h) =  \frac{1}{\ln(10)\sigma_{\rm log
    M_{*}}\sqrt{2\pi}} \times\hspace{0.65\columnwidth}
 \end{displaymath}
\begin{equation}
 \label{lognormal_equation}
\quad  \exp\left[-\frac{\left[\log_{10}(M_*)-\log_{10}(f_{\textsc{shmr}}(M_h))\right]^2}{2\sigma_{\rm log
    M_{*}}^2}\right]
\end{equation}

\noindent where $f_{\textsc{shmr}}$ represents the logarithmic mean of
the stellar mass given the halo mass for the $\Phi_c$ distribution
function. Equation \ref{lognormal_equation} is normalized such that
the integral of $\Phi_c(M_{*}|M_h)$ over $M_*$ is equal to 1.

To model $\Phi_c$ we must specify a functional form for both
$f_{\textsc{shmr}}$ and for $\sigma_{\rm log M_{*}}$. There is
increasing evidence to suggest that low and high mass galaxies have
different stellar-to-halo mass ratios, probably as a result of
multiple feedback mechanisms that operate at distinct mass scales and
regulate star formation. We therefore require a SHMR that is flexible
enough to capture such variations. We adopt the functional form
presented in \citet{Behroozi:2010} (hereafter ``B10'') which has been
shown to reproduce the local SDSS stellar mass function using the
abundance matching technique. In practice, $f_{\textsc{shmr}}(M_h)$ is
mathematically defined following via its inverse function:

\begin{displaymath}
\log_{10}(f_{\textsc{shmr}}^{-1}(M_\ast)) =  \log_{10}(M_h) = \hspace{0.65\columnwidth}
 \end{displaymath}
 \vspace{-3ex}
\begin{equation}
 \label{shmr}
\quad \log_{10}(M_1) + \beta\,\log_{10}\left(\frac{M_\ast}{M_{\ast,0}}\right) +
 \frac{\left(\frac{M_\ast}{M_{\ast,0}}\right)^\delta}{1 + \left(\frac{M_{\ast}}{M_{\ast,0}}\right)^{-\gamma}} - \frac{1}{2}
\end{equation}

\noindent where $M_{1}$ is a characteristic halo mass, $M_{*,0}$ is a
characteristic stellar mass, $\beta$ is the low mass end slope,
$\gamma$ controls the transition region, and $\delta$ controls the
massive end slope. Details regarding the justification of this
functional form can be found in $\S$ 3.4.3 of B10.

Note that a variety of similar functional forms have been proposed by
previous authors. For example, the interested reader can look at
Equation 2 in \citet{Moster:2010} and Equation 20 in
\citet{Yang:2009}.

In contrast to B10, we do not parametrize the redshift evolution of
this functional form. Instead, in Paper II, we bin the data into three
redshift bins and check for redshift evolution in the parameters a
posteriori. Another difference with respect to B10 is that we assume
that Equation \ref{shmr} is only relevant for central galaxies whereas
B10 assume that the SHMR also applies to satellite galaxies, provided
that the halo mass of a satellite galaxy is defined as M$_{\rm acc}$.

It is important to note that our SHMR traces the location of the
mean-log stellar mass: $f_{\textsc{shmr}}(M_h)\equiv \langle
\log_{10}(M_*(M_h))\rangle$. Other authors may report the mean stellar
mass, $\langle M_*(M_h)\rangle$, or even the mean halo mass at fixed
stellar mass, $\langle M_h(M_*)\rangle$
\citep[e.g.,][]{Conroy:2007}. These averaging systems will yield
different results in the presence of scatter. For example, $\langle
M_h(M_*)\rangle$ will be biased low compared to $\langle
\log_{10}(M_*(M_h))\rangle$ if $\sigma_{\log M_*}$ is non zero. This
bias will increase with $\sigma_{\log M_*}$ and for larger values of
the high-mass slope of $f_{\textsc{shmr}}^{-1}$.

In Figure \ref{mh_ms_vary_p} we illustrate the impact of the five
parameters that determine $f_{\textsc{shmr}}$ on the shape on the
SHMR. A brief description is as follows:

\begin{figure*}[htb]
\epsscale{1.23}
\plotone{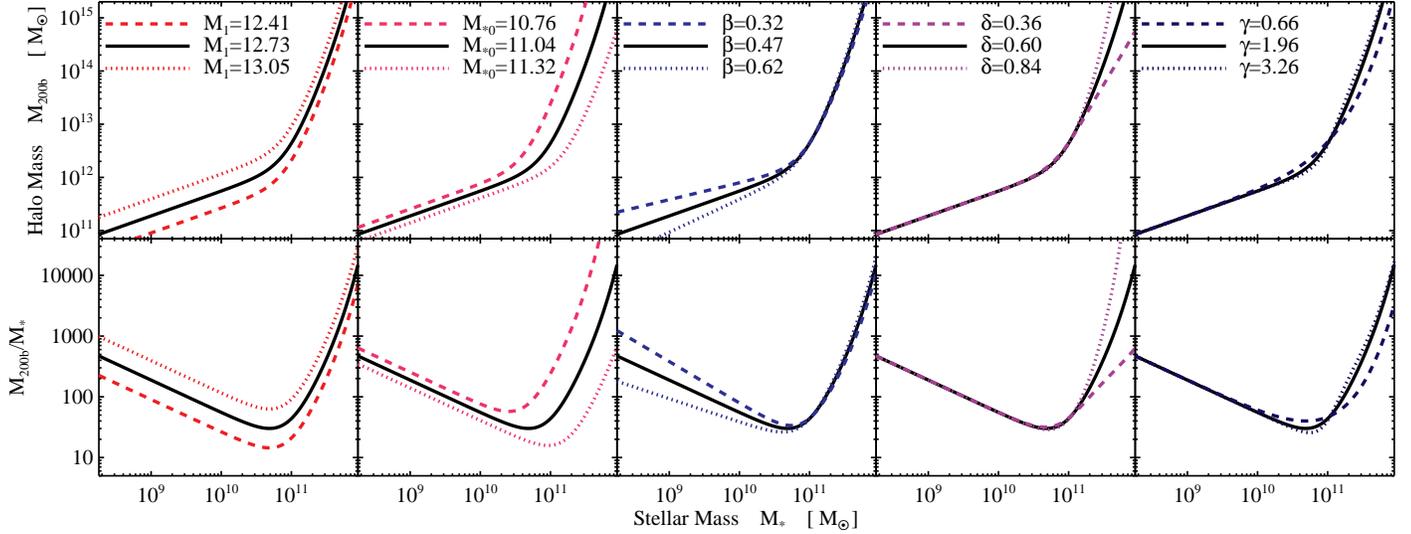}
\caption{Illustration of the influence of $M_{1}$, $M_{*,0}$, $\beta$,
  $\delta$ and $\gamma$ on the shape of the SHMR. $M_{1}$ controls the
  characteristic halo mass. $M_{*,0}$ controls the characteristic
  stellar mass. $\beta$ controls the low mass power-law
  slope. $\delta$ regulates how rapidly the SHMR climbs at high
  $M_*$. $\gamma$ controls the transition regime between the low-mass
  power-law regime and the high-mass sub-exponential behaviour.}
\label{mh_ms_vary_p}
\end{figure*}

\begin{itemize}
\item {\bf $M_{1}$} controls the characteristic halo mass; increasing
  $M_1$ will result in larger halos hosting galaxies at a given
  stellar mass. In Figure \ref{mh_ms_vary_p} (which represents $M_*$
  on the x-axis and $M_h$ on the y-axis): $M_1$ controls the y-axis
  amplitude (halo mass) of $f_{\textsc{shmr}}^{-1}$ so that a constant
  change in $M_{1}$ leads to a constant up/down logarithmic shift in
  $f_{\textsc{shmr}}^{-1}$. Note that this constant logarithmic shift
  may not be visually obvious because the slope of the SHMR increases
  sharply at $M_*>10^{11}$ M$_{\odot}$.
\item $M_{*,0}$ controls the characteristic stellar mass; increasing
  $M_{*,0}$ will result in smaller halos hosting galaxies at a given
  stellar mass. In Figure \ref{mh_ms_vary_p}, $M_{*,0}$ controls the
  x-axis amplitude of $f_{\textsc{shmr}}^{-1}$.
\item $\beta$ controls the low mass power-law slope of
  $f_{\textsc{shmr}}^{-1}$. When $\beta$ increases, the low mass end
  slope becomes steeper.
\item The $\delta$ parameter regulates how rapidly
  $f_{\textsc{shmr}}^{-1}$ climbs at high $M_*$. Indeed,
  $f_{\textsc{shmr}}^{-1}$ asymptotes to a sub-exponential function at
  high $M_*$ which signifies that $f_{\textsc{shmr}}^{-1}$ climbs more
  rapidly than a power-law function but less rapidly than an
  exponential function (see discussion in B10).
\item $\gamma$ controls the transition regime between the low-mass
  power-law regime and the high-mass sub-exponential behaviour. A
  larger value of $\gamma$ corresponds to a more sharp transition
  between the two regimes.
\end{itemize}

A quantity that is of particular interest is the mass (we refer here
to both $M_*$ and $M_h$) at which the ratio $M_h/M_*$ reaches a
minimum. This minimum is of noteworthy importance for galaxy formation
models because it marks the mass at which the accumulated stellar
growth of the central galaxy has been the most efficient. In this
paper, and in subsequent papers, we will refer to the stellar mass,
halo mass, and ratio at which this minimum occurs as the ``pivot
stellar mass'', $M_{*}^{\rm piv}$, the ``pivot halo mass'',
$M_{h}^{\rm piv}$, and the ``pivot ratio'', $(M_h/M_*)^{\rm
  piv}$. Note that $M_{*}^{\rm piv}$ and $M_{h}^{\rm piv}$ are not
simply equal to $M_{1}$ and to $M_{*,0}$. Indeed, the mathematical
formulation of the SHMR is such that the pivot masses depend on all
five parameters. The three parameters that have the strongest effect
on the pivot masses are $M_{1}$, $M_{*,0}$, and $\gamma$. For example,
as can be seen in the right hand panel of Figure \ref{mh_ms_vary_p},
$M_{*}^{\rm piv}$ and $M_{h}^{\rm piv}$ are inversely proportional to
$\gamma$. To lesser extent, the two remaining parameters, $\beta$ and
$\delta$ also have a small influence on the pivot masses.

\subsection{Scatter between stellar and halo mass}\label{scatter_shmr}

We now turn our attention to the second component of
$\Phi_c(M_{*}|M_h)$ which is the scatter in stellar mass at fixed halo
mass, $\sigma_{\rm log M_{*}}$. The total measured scatter will have
two components: an intrinsic component (noted $\sigma_{\rm log
  M_{*}}^{\rm i}$) and a measurement error component due to redshift,
photometry, and modeling uncertainties in stellar mass measurements
(noted $\sigma_{\rm log M_{*}}^{\rm m}$). It is reasonable to assume
that the intrinsic scatter component is independent of the measurement
error component. Assuming Gaussian error distributions, we can write
that:

\begin{equation}
  (\sigma_{\rm log M_{*}})^2=(\sigma_{\rm log M_{*}}^{\rm i})^2+(\sigma_{\rm log M_{*}}^{\rm m})^2. 
\end{equation}

While the error distribution for the stellar mass estimate for any
single galaxy may be non Gaussian, in this work we are only concerned
with stacked ensembles. B10 have tested that the error distribution
for a stacked ensemble is Gaussian to good approximation and that
small non-Gaussian wings in this distribution are not likely to affect
this type of analysis. 

In practice, since the data are always binned according to $M_*$, the
observables are actually sensitive to the scatter in halo mass at
fixed stellar mass which we note $\sigma_{\rm log M_h}$. It will
therefore be useful to understand the link between $\sigma_{\rm log
  M_h}$ and $\sigma_{\rm log M_{*}}$. If the SHMR is a power law, the
relationship between $\sigma_{\rm log M_h}$ and $\sigma_{\rm log
  M_{*}}$ is simply:

\begin{equation}
  \sigma_{\rm log M_h}=\sigma_{\rm log M_{*}}\frac{{\rm d}(\log_{10} M_h)}{{\rm d}(\log_{10} M_* )}.
  \label{sigmas}
\end{equation}

For example, if there is a power law relation between halo mass and
stellar mass such that $M_h=M_*^\eta$ then $\sigma_{\rm log M_h}=\eta
\times \sigma_{\rm log M_{*}}$. In our case, the SHMR behaves like a
power-law at low $M_*$. At high $M_*$, however, ${\rm
  d}(\log_{10}M_h)/{\rm d}(\log_{10}M_*)$ increases as a function of
$M_*$. Therefore if $\sigma_{\rm log M_{*}}$ is constant, $\sigma_{\rm
  log M_h}$ will be equal to $\sigma_{\rm log M_h}=\beta \times
\sigma_{\rm log M_{*}}$ at low $M_*$ but then will continuously
increase with $M_*$ at a rate set by $\gamma$ and $\delta$.

If we adopt the best fit model to the SHMR from Paper II in the
redshift range $0.22<z<0.48$, we find that the power-law index of the
SHMR increases steeply at $\log_{10}(M_*)>11$ so that $\sigma_{\rm log
  M_h}$ becomes quite large. For example, $\sigma_{\rm log M_h}\sim
0.46$ dex at $\log_{10}(M_*)=11$ and $\sigma_{\rm log M_h}\sim 0.7$
dex at $\log_{10}(M_*)=11.5$. In practical terms, this implies that
the most massive galaxies do not necessarily live in the most massive
halos. For example, a galaxy with $M_* \sim 2\times 10^{11}$
M$_{\odot}$ could be the central galaxy of of group with $M_h \sim
10^{13}-10^{14}~{\rm M}_{\odot}$, or could also be the central galaxy
of a cluster with $M_h>10^{15}~{\rm M}_{\odot}$. The increase of
$\sigma_{\rm log M_h}$ with $M_*$ will lead to a noticeable effect in
the g-g lensing, clustering, and stellar mass functions at large $M_*$
that is analogous to Eddington bias. This effect will be discussed in
further detail in $\S$ \ref{parameter_influance}.


\section{HOD Framework}\label{hodtheory}

In this section, we show how $\Phi_c(M_{*}|M_h)$ can be used to
determine the central halo occupation function and we introduce five
new parameters to describe the satellite occupation function.

\subsection{Halo Occupation Functions}\label{thehodmodel}

In this paper, we assume that stellar mass is used to implement the
HOD model since it is expected to be a more faithful tracer of halo
mass than galaxy luminosity.

Consider a galaxy sample such that $M_*>M_{*}^{t_1}$ (a ``threshold''
sample). The {\it central occupation function}, noted $\langle N_{\rm
  cen}(M_h|M_{*}^{t_1})\rangle$, is the average number of central
galaxies in this sample that are hosted by a halo of mass $M_h$. The
{\it satellite occupation function}, noted $\langle N_{\rm
  sat}(M_h|M_{*}^{t_1})\rangle$, is the equivalent function for
satellite galaxies. 

In what follows, we focus on the appropriate equations for threshold
samples. In Paper II however, we will use ``binned'' samples
($M_{*}^{t_1}<M_*<M_{*}^{t_2}$) to calculate the g-g lensing and the
SMF. We therefore note that the occupation functions for binned
samples are trivially derived from the occupation function for
threshold samples via:

\begin{equation}
\langle N_{\rm cen}(M_h|M_{*}^{t_1},M_{*}^{t_2}) \rangle= \langle
N_{\rm cen}(M_h|M_{*}^{t_1})\rangle-\langle N_{\rm cen}(M_h|M_{*}^{t_2})\rangle,
\end{equation}

\noindent and

\begin{equation}
 \langle N_{\rm sat}(M_h|M_{*}^{t_1},M_{*}^{t_2}) \rangle= \langle
N_{\rm sat}(M_h|M_{*}^{t_1})\rangle-\langle N_{\rm sat}(M_h|M_{*}^{t_2})\rangle.
\end{equation}


\subsection{Functional form for $\ncen$}\label{ncen_model}

For a threshold sample of galaxies, $\langle N_{\rm
  cen}(M_h|M_{*}^{t_1})\rangle$ is fully specified given
$\Phi_c(M_{*}|M_h)$ according to:

\begin{equation}
\langle N_{\rm cen}(M_h|M_{*}^{t_1})\rangle = \int_{M_{*}^{t_1}}^{\infty} \Phi_c(M_*|M_h) {\rm d}M_* .
\label{ncen_and_phic}
\end{equation}

Because the integral of $\Phi_c(M_{*}|M_h)$ over $M_*$ is equal to 1,
$\langle N_{\rm cen}(M_h|M_{*}^{t_1})\rangle$ will vary between 0 and
1.

To begin with, let us make the simplifying assumption that
$\sigma_{\rm log M_{*}}$ is constant. Because $\Phi_c$ is parametrized
as a log-normal distribution, the central occupation function can be
analytically derived from Equation \ref{ncen_and_phic} by considering
the cumulative distribution function of the Gaussian:

\begin{displaymath}
  \langle N_{\rm cen}(M_h|M_{*}^{t_1}) \rangle = \hspace{0.65\columnwidth}
\end{displaymath}
\begin{equation}
\label{ncen}
\frac{1}{2}\left[ 1-\mbox{erf}\left(\frac{\log_{10}(M_{*}^{t_1}) - \log_{10}(f_{\textsc{shmr}}(M_h)) }{\sqrt{2}\sigma_{\rm log M_{*}}} \right)\right],
\end{equation}

\noindent where erf is the error function defined as:

\begin{equation}
\label{erf_funct}
\mbox{erf}(x)=\frac{2}{\sqrt{\pi}}\int_0^{x}e^{-t^2} {\rm d}t.
\end{equation}

It is important to note that Equation \ref{ncen} is only valid when
$\sigma_{\rm log M_{*}}$ is constant. In the more general case where
$\sigma_{\rm log M_{*}}$ varies, $\ncen$ can nonetheless be calculated
by numerically integrating Equation \ref{ncen_and_phic}. In Paper II
we will consider cases in which $\sigma_{\rm log M_{*}}$ varies due to
the effects of stellar mass dependant measurement errors. In this
case, we will numerically integrate Equation \ref{ncen_and_phic} to
calculate $\ncen$ (see $\S$ 4.2 in Paper II).

We note that most readers may be more familiar with a simplified
version of Equation \ref{ncen} that assumes that
$f_{\textsc{shmr}}(M_h)$ is a power law. We will now describe the
assumptions made in order to obtain the more commonly employed
equation for $\ncen$ from Equation \ref{ncen}.

If we make the assumption that $f_{\textsc{shmr}}(M_h) \propto M_h^p$
and we define $M_{\rm min}$ such that $M_{\rm min}\equiv
f_{\textsc{shmr}}^{-1}(M_{*}^{t_1})$ (in other terms, $M_{\rm min}$ is
the inverse of the SHMR relation for the stellar mass threshold
$M_{*}^{t_1}$) then using Equation \ref{ncen} we can write that:

\begin{displaymath}
  \langle N_{\rm cen}(M_h|M_{*}^{t_1}) \rangle = \hspace{0.65\columnwidth}
\end{displaymath}
\begin{equation}
\frac{1}{2}\left[ 1-\mbox{erf}\left(\frac{\log_{10}(M_{\rm min}^p) - \log_{10}(M_h^p) }{\sqrt{2}\sigma_{\rm log M_{*}}} \right)\right].
\end{equation}

If we now use the fact that $\rm{erf}(-x)= -\rm{erf}(x)$ and if we
{\em define} $\widetilde{\sigma}_{\rm{log}M}$ such that
$\widetilde{\sigma}_{\rm{log}M}\equiv \sigma_{\rm log M_{*}}/p$ we can
write that:

\begin{displaymath}
\langle N_{\rm cen}(M_h|M_{*}^{t_1}) \rangle  = \hspace{0.65\columnwidth}
\end{displaymath}
\begin{equation}
\frac{1}{2}\left[ 1+\mbox{erf}\left(\frac{\log_{10}(M_h) - \log_{10}(M_{\rm min}) }{\sqrt{2}\widetilde{\sigma}_{\rm{log}M}} \right)\right],
\label{ncen2}
\end{equation}

\noindent which is a commonly employed formula for $\ncen$. Firstly,
it is important to note that Equation \ref{ncen2} is only an
approximation for $\ncen$ for the case when the SHMR is a power-law
and is certainly not valid over a large range of stellar
masses. Secondly, $\widetilde{\sigma}_{\rm{log}M}$ can be interpreted
as the scatter in halo mass at fixed stellar mass if and only if the
SHMR is a power-law and if $\sigma_{\rm log M_{*}}$ is constant. Since
there is accumulating evidence that the SHMR is not a single power law
(and the same is in general true for the relationship between halo
mass and galaxy luminosity), we recommend using Equation \ref{ncen}
instead of Equation \ref{ncen2}.

\begin{figure*}[htb]
\epsscale{1.22} 
\plotone{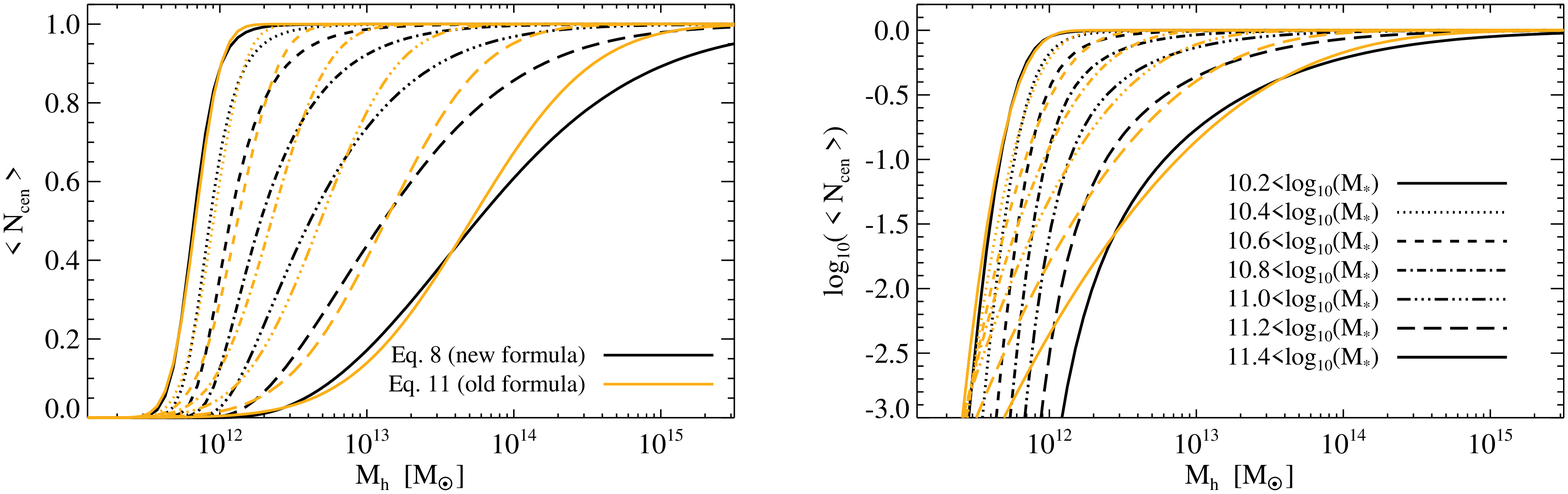}
\caption{Impact of the analytic model for the mean number of central
  galaxies in a given halo.  Black lines show the form of $\ncen$ for
  stellar mass threshold samples using Equation \ref{ncen}. Orange
  lines show $\ncen$ when Equation \ref{ncen2} is used to describe the
  clustering instead of Equation \ref{ncen}. The parametric form of
  the SHMR is a power law at low $M_*$ and thus Equation \ref{ncen2}
  provides a reasonable description of $\ncen$ at
  $\log_{10}(M_*)\lesssim 10.5$. At $\log_{10}(M_*)\gtrsim 10.5$
  however, $\ncen$ deviates from a simple erf function and there are
  noticeable differences between the two proposed forms for $\ncen$.}
\label{study_mmin}
\end{figure*}

\begin{figure}[htb]
\epsscale{1.25}
\plotone{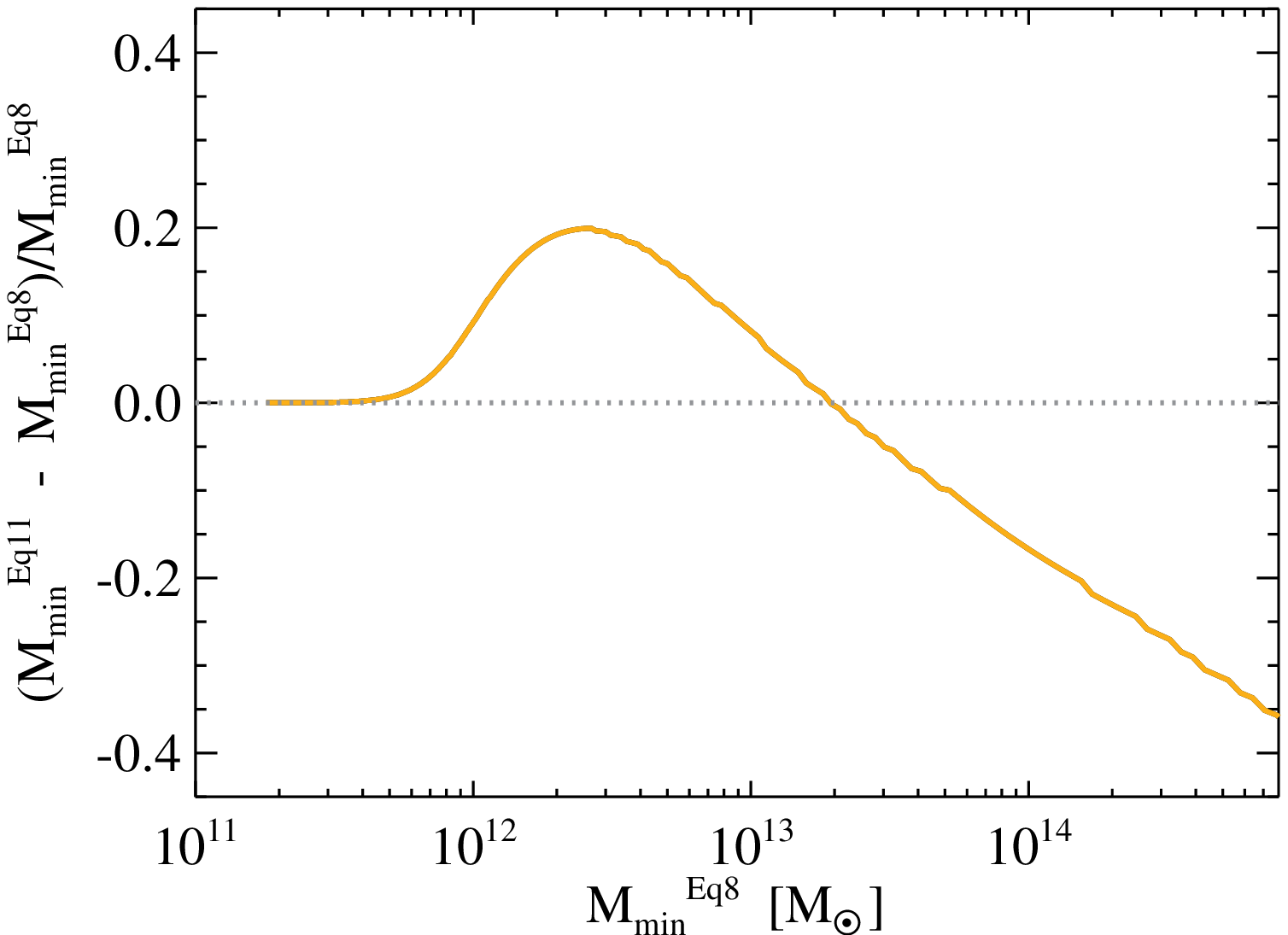}
\caption{Difference between two analytic models for the mean number of
  central galaxies in a given halo.  Assuming that Equation \ref{ncen}
  correctly represents $\ncen$, we evaluate the difference between the
  value for $M_{\rm min}$ if Equation \ref{ncen2} is used (noted here
  as $M_{\rm min}^{Eq11}$) to fit clustering data and
  $f_{\textsc{shmr}}^{-1}(M_{*}^{t_1})$ (noted here as $M_{\rm
    min}^{Eq8}$). The difference is negligible below $\log_{10}(M_{\rm
    min})\lesssim 12$ but can be of order 10 to 40\% at higher
  masses.}
\label{study_mmin2}
\end{figure}

Figures \ref{study_mmin} and \ref{study_mmin2} illustrate the
difference in $\ncen$ when Equation \ref{ncen} is used to describe
clustering instead of Equation \ref{ncen2}. To make this figure, we
have assumed the parameter set: $\log_{10}(M_{1})= 12.71$,
$\log_{10}(M_{*,0})= 11.04$, $\beta= 0.467$, $\delta= 0.62$, $\gamma=
1.89$, and $\sigma_{\rm log M_{*}}$=0.25. For each stellar mass
threshold in Figure \ref{study_mmin}, we determine the values of
$\mmin$ and $\widetilde{\sigma}_{\rm{log}M}$ in Equation \ref{ncen2}
such that the number density of central galaxies and the bias of those
galaxies is the same as that achieved with Equation \ref{ncen}. Thus
our procedure mimics what one would obtain through analysis of the
clustering and space density of such samples, assuming that the
satellite occupation would be the same in either analysis.

Figure \ref{study_mmin} reveals that because the SHMR has a
sub-exponential behaviour at $\log_{10}(M_*)\gtrsim 10.5$, $\ncen$
begins to deviate from a simple erf function for high stellar mass
samples and therefore is not well described by Equation
\ref{ncen2}. Assuming that Equation \ref{ncen} correctly represents
$\ncen$, the error made on $M_{\rm min}$ can be of order 10\% to 40\% at
$M_h=f_{\textsc{shmr}}^{-1}(M_{*}^{t_1})\gtrsim 10^{12}$ $M_{\odot}$ if
Equation \ref{ncen2} is used to describe $\ncen$ instead of Equation
\ref{ncen}.

We note that this does not invalidate Equation \ref{ncen2} as a
possible parameterization of the central occupation function. However,
interpreting $M_{\rm min}$ in Equation \ref{ncen2} as
$f_{\textsc{shmr}}^{-1}(M_{*}^{t_1})$ can result in a 10-40\% error in
the true mean halo mass (with larger errors for $\sigma_{\rm log
  M_{*}}>0.25$). Also, the ``scatter''
($\widetilde{\sigma}_{\rm{log}M}$) constrained by this parametrization
is not equal to the scatter in a log-normal distribution of stellar
mass at fixed halo mass. Finally, one troublesome aspect of using the
erf functional form in Equation \ref{ncen2} is that $\ncen$ curves for
different stellar mass thresholds may actually cross at low halo mass,
implying the unphysical condition that halos of mass $M_h$ have a
``negative'' amount of galaxies between two threshold values.  This is
seen at $M_h\sim 10^{11.5}$ M$_{\odot}$ in Figure
\ref{study_mmin}. For values of $\sigma_{\rm log M_{*}}$ larger than
what we have assumed here, this effect will occur at even higher halo
mass. Using Equation \ref{ncen} with a model for the SHMR prevents
this from occurring, and $\ncen$ for various galaxy samples can be
calculated self-consistently.

\subsection{Functional form for $\nsat$ }\label{model_nsat}

In addition to the five parameters introduced to model $\ncen$ and
$\sigma_{\rm log M_{*}}$, we introduce five new parameters to model
$\nsat$. In order to simultaneously fit g-g lensing, clustering, and
stellar mass function measurements that employ different binning
schemes, we require a model for $\nsat$ that is independent of any
given binning scheme. For this reason, we parameterize the satellite
function with threshold samples. The number of satellite galaxies in a
bin of stellar mass is determined by a simple difference of two
threshold samples. This also eliminates the need to integrate over
stellar mass, as required when working explicitly through the
conditional stellar mass function.

Numerical simulations demonstrate that the occupation of sub-halos
(e.g., \citealt{Kravtsov:2004,Conroy:2006}) and satellite galaxies in
cosmological hydrodynamic simulations (\citealt{Zheng:2005}) follow a
power law at high host halo mass, then fall off rapidly when the mean
occupation becomes significantly less than unity. Thus we parametrize
the satellite occupation function as a power of host mass with an
exponential cutoff and scaled to $\ncen$ as follows:

\begin{displaymath}
  \langle N_{\rm sat}(M_h|M_{*}^{t_1}) \rangle = \hspace{0.65\columnwidth}
\end{displaymath}
\begin{equation}
\label{e.nsat}
\langle N_{\rm cen}(M_h|M_{*}^{t_1}) \rangle \left(\frac{M_h}{M_{\rm sat}}\right)^{\asat} \exp\left(\frac{-M_{\rm cut}}{M_h}\right),
\end{equation}

\noindent where $\alpha_{\rm sat}$ represents the power-law slope of
the satellite mean occupation function, $M_{\rm sat}$ defines the
amplitude of the power-law, and $M_{\rm cut}$ sets the scale of the
exponential cut-off. Here, $M_h$ refers to the host halo mass of
  satellite galaxies.

Observational analyses have demonstrated that there is a
self-similarity in occupation functions such that $M_{\rm sat}/M_{\rm
  min}\approx constant$ for luminosity-defined samples
(\citealt{Zehavi:2005,Zheng:2007, Tinker:2007, Zheng:2009,
  Zehavi:2010,Abbas:2010}), where $M_{\rm min}$ is taken from Equation
\ref{ncen2} and is conceptually similar to
$f_{\textsc{shmr}}^{-1}(M_{*}^{t_1})$ (modulo a $10-40\%$ difference
as shown in Figure \ref{study_mmin2}) and where $M_{*}^{t_1}$ is the
stellar mass threshold of the sample. Instead of simply modelling
$M_{\rm sat}$ and $M_{\rm cut}$ as constant factors of $\fshmr$, we
add flexibility to our model by enabling $M_{\rm sat}$ and $M_{\rm
  cut}$ to vary as power law functions of $\fshmr$:

\begin{equation}
\frac{M_{\rm sat}}{10^{12} M_{\odot}}= B_{\rm sat} \left(\frac{\fshmr}{10^{12} M_{\odot}}\right)^{\beta_{\rm sat}},
\end{equation}

and 

\begin{equation}
\label{mcut_eq}
\frac{M_{\rm cut}}{10^{12} M_{\odot}}= B_{\rm cut} \left(\frac{\fshmr}{10^{12} M_{\odot}}\right)^{\beta_{\rm cut}}.
\end{equation}

\citet{Zheng:2007} find that $M_{\rm sat}/M_{\rm min} \sim 18$ for
SDSS and $M_{\rm sat}/M_{\rm min} \sim 16$ using luminosity defined
samples in DEEP2. For $B_{\rm cut}$, the expectation is that the
cutoff mass scale occurs at $M_{\rm min}<M_{\rm cut}<M_{\rm sat}$,
although it can be significantly smaller.

\subsection{Total stellar mass as a function of halo mass}

Using this model, one can also compute the total amount of stellar
mass in galaxies (summing the contribution from both centrals and
satellites) as a function of halo mass $M_h$. To begin with, let us
consider the total stellar mass as a function of halo mass in some
stellar mass bin: $M_{*}^{\rm tot}(M_h|M_*^{t1},M_*^{t2})$. The
expression for $M_{*}^{\rm tot}(M_h|M_*^{t1},M_*^{t2})$ is given by:

\begin{displaymath}
M_{*}^{\rm tot}(M_h|M_*^{t1},M_*^{t2}) = \hspace{0.65\columnwidth}
\end{displaymath}
\begin{equation}
\label{tot_msat_1}
\int_{M_*^{t1}}^{M_*^{t2}}\left[\Phi_c(M_*|M_h)+\Phi_s(M_*|M_h)\right]M_*{\rm d}M_* .
\end{equation}

However, in the previous section we have only specified the analytic
form for $\Phi_c$ (Equation \ref{lognormal_equation}) but not for
$\Phi_s$. Indeed, in $\S$ \ref{model_nsat} we outlined an analytic
model for $\nsat$ but calculating the analytic derivative of $\nsat$
would be tedious. Thankfully, however, we do not specifically need to
know the functional form of $\Phi_s$ in order to calculate Equation
\ref{tot_msat_1}. By using the integration by parts rule, we can
re-write Equation \ref{tot_msat_1} in a more convenient form as
follows:

\begin{displaymath}
M_{*}^{\rm tot}(M_h|M_*^{t1},M_*^{t2}) = \hspace{0.65\columnwidth}
\end{displaymath}
\begin{eqnarray}
&& \int_{M_*^{t1}}^{M_*^{t2}} \langle N_{\rm cen}(M_h|M_{*}) \rangle {\rm d}M_* - \left[ \langle N_{\rm cen}(M_h|M_{*}) \rangle M_*\right]_{M_*^{t1}}^{M_*^{t2}} \nonumber\\
&+&\int_{M_*^{t1}}^{M_*^{t2}} \langle N_{\rm sat}(M_h|M_{*}) \rangle
{\rm d}M_* - \left[ \langle N_{\rm sat}(M_h|M_{*}) \rangle
  M_*\right]_{M_*^{t1}}^{M_*^{t2}}.  \nonumber\\
&&
\label{tot_msat_2}
\end{eqnarray}

This equation provides us with a convenient way to calculate the total
stellar mass locked up in galaxies with $M_*^{t1}<M_*<M_*^{t2}$ as a
function of halo mass $M_h$.

\subsection{Summary of model parameters}

\begin{figure*}[htb]
\epsscale{1.0} 
\plotone{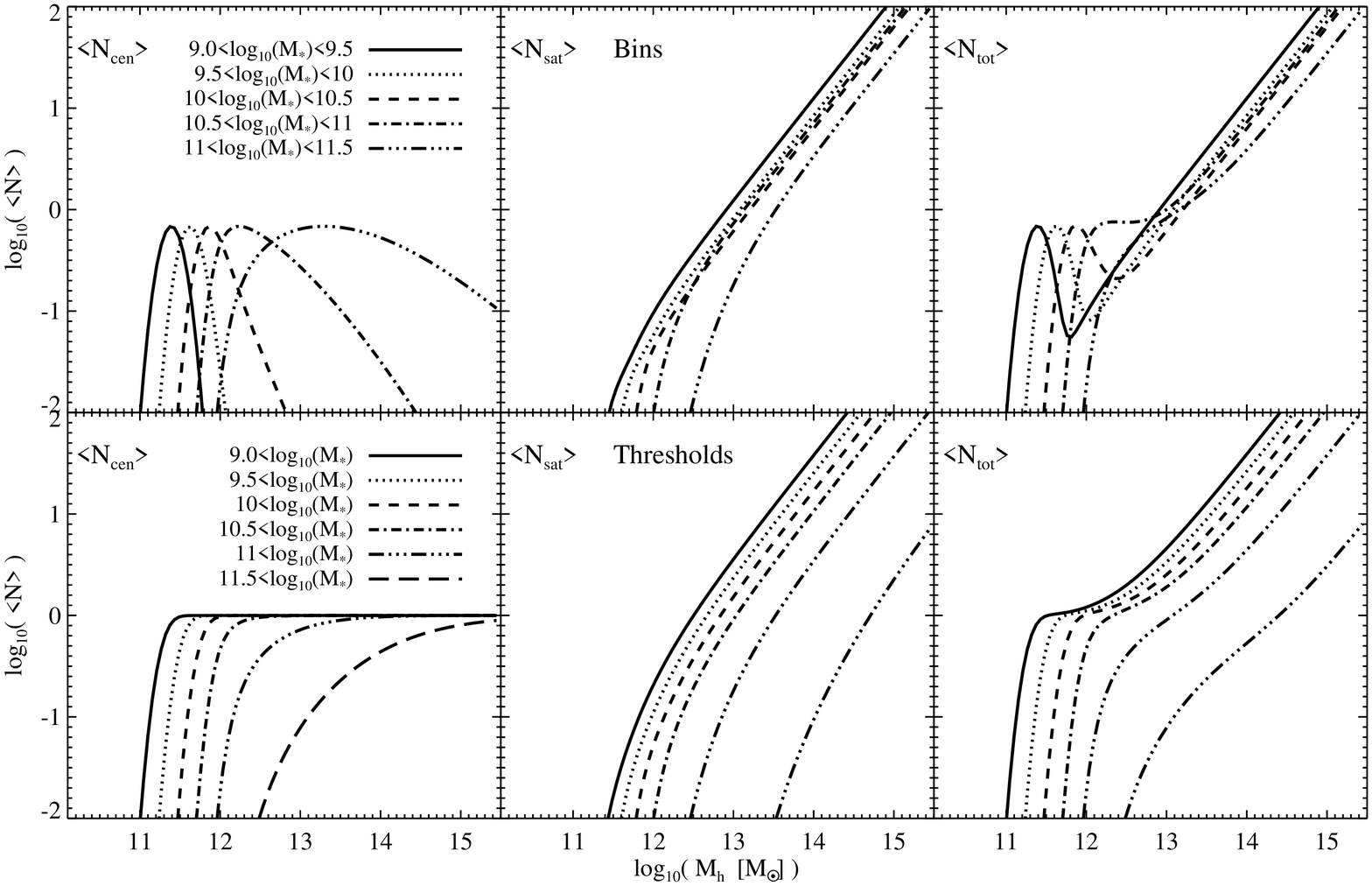}
\caption{Illustration of the occupation functions for various galaxy
  samples as a function of stellar mass. The upper panels represent
  binned galaxy samples whereas the lower panel represent threshold
  samples. Left panels: central occupation function. Middle panels:
  satellite occupation function. Right panels: total occupation
  function where $\ntot=\ncen+\nsat$. The parameters chosen for this
  HOD model correspond to the best fit parameters from Paper II for
  $0.48<z<0.74$.}
\label{hod_bins_illustration}
\end{figure*}

In total, we have introduced six parameters to model the central
occupation function ($M_{1}$, $M_{*,0}$, $\beta$, $\delta$, $\gamma$,
$\sigma_{\rm log M_{*}}$) and five parameters to model the satellite
occupation function ($\beta_{\rm sat}$, $B_{\rm sat}$, $\beta_{\rm
  cut}$, $B_{\rm cut}$, $\alpha$). In addition, one could introduce a
model for $\sigma_{\rm log M_{*}}$ or assume that the scatter is
constant in which case there would be a total of eleven parameters for
this model. In Figure 8 of Paper II, we show the two dimensional
marginalized distributions for this parameter set using data from the
COSMOS survey. The model described in this paper provides an excellent
fit to COSMOS data. Figure 8 in Paper II demonstrates that this model
is reasonably free of parameter degeneracies. A summary and
description of these parameters can be found in Table
\ref{free_params}.  Figure \ref{hod_bins_illustration} gives an
illustration of the central and satellite occupation functions for
galaxy samples in bins and thresholds of stellar mass.

\begin{deluxetable*}{llll} 
\tablecolumns{4} \tablecaption{Parameters in model\label{free_params}} \tablewidth{0pt} 
\startdata
\hline 
\hline 
\\  [-1.5ex]
Parameter & Unit & Description & $\ncen$ or $\nsat$ \\ [1ex]
\hline\\[-1.5ex]
$M_{1}$   & $M_{\sun}$   & Characteristic halo mass in the SHMR &$\ncen$ \\ 
$M_{*,0}$  & $M_{\sun}$  & Characteristic stellar mass in the SHMR &$\ncen$\\ 
$\beta$   & none & Faint end slope in the SHMR &$\ncen$ \\
$\delta$  & none & Controls massive end slope in the SHMR &$\ncen$ \\
$\gamma$  & none & Controls the transition regime in the SHMR &$\ncen$ \\
$\sigma_{\rm log M_{*}}$ & dex    & Log-normal scatter in stellar mass at fixed halo mass&$\ncen$ \\
$\beta_{\rm sat}$ & none   & Slope of the scaling of $M_{\rm sat}$ &$\nsat$ \\
$B_{\rm sat}$  & none & Normalization of the scaling of $M_{\rm sat}$  &$\nsat$ \\
$\beta_{\rm cut}$ & none   & Slope of the scaling of $M_{\rm cut}$ &$\nsat$ \\
$B_{\rm cut}$  & none & Normalization of the scaling of $M_{\rm cut}$ &$\nsat$ \\
$\alpha_{\rm sat}$ & none    & Power-law slope of the satellite occupation function&$\nsat$ \\
\enddata
\end{deluxetable*}

\section{How to derive the SMF, g-g lensing, and clustering from the
  model}\label{application_of_model}

We now describe how the model outlined in the previous section yields
analytic descriptions for the g-g lensing, clustering, and stellar
mass function which can then be fit simultaneously to observations.

\subsection{Analytical  model for the stellar mass function}

The stellar mass function is typically calculated in bins in stellar
mass. Let us consider the stellar mass bin
$M_*^{t_1}<M_*<M_*^{t_2}$. The abundance of galaxies within this
stellar mass bin, $\Phi_{SMF}(M_*^{t_1},M_*^{t_2})$, is simply
obtained from our model and the halo mass function, ${\rm
  d}n/{\rm{d}}M_h$, according to:

\begin{displaymath}
\Phi_{SMF}(M_*^{t_1},M_*^{t_2}) \hspace{0.65\columnwidth}
\end{displaymath}
\begin{eqnarray}
&=&\int_{0}^{\infty} \left[\int_{M_*^{t_1}}^{M_*^{t_2}}
  \Phi(M_{*}|M_h) {\rm{d}}M_* \right] \frac{{\rm{d}}n}{{\rm{d}}M_h}{\rm{d}}M_h\nonumber\\
&=&\int_0^{\infty} \ntotb
 \frac{{\rm{d}}n}{{\rm{d}}M_h}{\rm{d}}M_h,
\end{eqnarray}

\noindent where we recall that $\Phi(M_{*}|M_h)$ represents the
conditional stellar mass function and $\langle N_{\rm tot}\rangle$ is
the total occupation function (including both satellites and
centrals).

\subsection{The Lensing Observable, $\Delta\Sigma$}

The shear signal induced by a given foreground mass distribution on a
background source galaxy will depend on the transverse proper distance
between the lens and the source and on the redshift configuration of
the lens-source system. A lens with a projected surface mass density,
$\Sigma(r)$, will create a shear that is proportional to the {\em
  surface mass density contrast}, $\Delta\Sigma(r)$:

\begin{equation}
  \Delta \Sigma(r)\equiv\overline{\Sigma}(< r)-\overline{\Sigma}(r)=\Sigma_{\rm crit}\times\gamma_t(r).
\label{dsigma}
\end{equation}

Here, $\overline{\Sigma}(< r)$ is the mean surface density within
proper radius $r$, $\overline{\Sigma}(r)$ is the azimuthally averaged
surface density at radius $r$
\citep[e.g.,][]{Miralda-Escude:1991,Wilson:2001}, and $\gamma_t$ is
the tangentially projected shear. The geometry of the lens-source
system intervenes through the {\em critical surface mass
  density}\footnotemark[2]\footnotetext[2]{Note that some authors
  consider the comoving critical surface mass density which has an
  extra factor of $(1+z)^{-2}$ with respect to ours.},
$\Sigma_\mathrm{crit}$, which depends on the angular diameter
distances to the lens ($D_{\rm OL}$), to the source ($D_{\rm OS}$),
and between the lens and source ($D_{\rm LS}$):

\begin{equation}
  \Sigma_\mathrm{crit} = \frac{c^2}{4\pi G_{{{\rm N}}}}\,
  \frac{D_\mathrm{OS}}{D_\mathrm{OL}\,D_\mathrm{LS}}\;,
\label{sigma_crit}
\end{equation}

\noindent where $G_{\rm N}$ represents Newton's constant. 


\subsection{Relationship between $\Delta\Sigma$, the density field,
  and correlation functions}\label{ds_correlation_funct}

Consider two different populations characterised respectively by
$\delta_a$ and $\delta_b$. The {\em two-point cross-correlation
  function} of $\delta_a$ and $\delta_b$ at comoving position $\vec
r_{\rm co}$ is given by:

\begin{equation}
\xi_{ab}(\vec r_{\rm co}) \equiv \langle \delta_a(\vec r_{\rm
  co})\delta_{b}(\vec x_{\rm co}+\vec r_{\rm co}) \rangle .
\end{equation}

For example, if $\delta_{\rm g}$ and $\delta_{\rm dm}$ are
respectively the over-densities of galaxies and dark matter, then we
can characterize their relative distributions via the {\em galaxy-mass
  cross-correlation function} which is noted $\xi_{\rm gm}$ and is
equal to:

\begin{equation}
  \xi_{\rm gm}(\vec{r_{\rm co}}) = \langle \delta_{\rm g} (\vec{r_{\rm co}})
  \delta_{\rm dm}(\vec{x_{\rm co}}+\vec{r_{\rm co}})\rangle . 
\end{equation}

Similarly, $\xi_{\rm gg}$ refers to the {\em galaxy auto-correlation
  function}.

In the following, $\vec r_{\rm co}$ is the three dimensional comoving
distance, $\vec r_{||,{\rm co}}$ is the projected comoving line-of
sight distance, and $\vec r_{p,{\rm co}}$ is the projected comoving
transverse distance:

\begin{equation}
 r_{\rm co}=\sqrt{r_{p,{\rm co}}^2+r_{||,{\rm co}}^2}.
\label{r_p_co}
\end{equation}

In Paper II, we will employ physical coordinates for g-g lensing
measurements whereas for clustering we will use comoving
coordinates. In previous work, \citet{Mandelbaum:2006c} have used
comoving coordinates and \citet{Johnston:2007} have used physical
coordinates. Therefore, our g-g lensing formulas more closely resemble
those of \citet{Johnston:2007}. The relationship between comoving and
physical distances is simply $r_{\rm co}=r_{\rm ph} (1+z)$. In a
similar fashion to Equation \ref{r_p_co}, we can write that:

\begin{equation}
 r_{\rm ph}=\sqrt{r_{p,{\rm ph}}^2+r_{||,{\rm ph}}^2}.
\label{r_p_ph}
\end{equation}

In comoving coordinates, the density field is $\rho(r_{\rm
  co},z)=\overline \rho(1+\xi_{\rm gm}(r_{\rm co},z))$ where
$\overline \rho=\rho_{c,0} \Omega_{m,0}$ is the average density of
matter in the Universe. Since $\xi_{\rm gm}$ is often expressed in
comoving coordinates, we first derive $\Sigma$ in comoving units
(noted $\Sigma_{\rm co}$) and then we transform $\Sigma$ into physical
units (noted $\Sigma_{\rm ph}$) before computing $\Delta\Sigma$. For a
lens at redshift $z_L$, the projected surface mass density,
$\Sigma$, is obtained by integrating the 3d density over the
line-of-sight:

\begin{displaymath}
  \Sigma_{\rm co}(r_{\rm p,co},z_L) \hspace{0.65\columnwidth}
\end{displaymath}
\begin{eqnarray}
&=& \int \rho\left(\sqrt{r_{\rm p,co}^2+r_{||,\rm co}^2},z_L\right){\rm d}r_{\rm ||,co} \nonumber\\
&=& \rho_{c,0} \Omega_{m,0} \int \left[1+\xi_{\rm
    gm}\left(\sqrt{r_{\rm p,co}^2+r_{||,\rm
        co}^2},z_L\right)\right]{\rm d}r_{\rm ||,co}, \nonumber\\
&&
\label{sig_and_rho}
\end{eqnarray}

\noindent where $r_{\rm p,co}$ and $r_{\rm ||,co}$ refer respectively
to the comoving transverse and line-of-sight distance from the
lens. In principle, this integral should extend from the redshift of
the observer ($z_O$) to the redshift of the source ($z_S$). However,
\cgm\ falls off rapidly enough that in practice, the redshift
evolution of \cgm\ can be neglected and integrating only out to a
distance of $r_{\rm ||,co}=50$ Mpc is sufficient. Furthermore, for the
purpose of computing $\Delta\Sigma$, the constant term in Equation
\ref{sig_and_rho} can be dropped (due to the subtraction in
$\Delta\Sigma$) and the mean excess projected density
$\overline\Sigma(r)$ can be approximated by the radial integral:

\begin{displaymath}
\overline\Sigma_{\rm co}(r_{\rm p,co},z_L)   = \hspace{0.65\columnwidth}
\end{displaymath}
\begin{equation}
 2 \rho_{c,0} \Omega_{m,0} \int_0^{50 {\rm Mpc}}\xi_{\rm gm}\left(\sqrt{r_{\rm p,co}^2+r_{||,\rm co}^2},z_L\right){\rm d}r_{\rm ||,co}.
\label{dsigma1_bis}
\end{equation}

The mean excess projected density is physical units is then
$\overline\Sigma_{\rm ph} =\overline\Sigma_{\rm co}\times(1+z)^2$.

The average $\overline\Sigma$ within radius $r$ is equal to:

\begin{equation}
\overline\Sigma_{\rm ph}(<r_{\rm p,ph})  = \frac{2}{r_{\rm p,ph}^2} 
\int_0^{r_{\rm p,ph}}\overline \Sigma_{\rm ph}(r^\prime)r^\prime   {\rm
  d}r^\prime .
\label{dsigma2}
\end{equation}

Finally, $\Delta\Sigma$ is obtained by combining Equations
\ref{dsigma}, \ref{dsigma1_bis} and \ref{dsigma2}.

\subsection{Analytical modeling of $\xi_{gg}$ and $w(\theta)$}\label{model_for_xi_gg}

Our model for calculating the autocorrelation function of galaxies is
based on the model given in \cite{Tinker:2005} \citep[also
  see][]{Zheng:2004}. As described above, the HOD is broken into
central and satellite galaxy occupation functions. Thus, in the HOD
context, pairs of galaxies come from two distinct terms; pairs within
a single halo and pairs between galaxies in two different halos. The
total correlation function is

\begin{equation}
\xi_{gg}(r_{\rm co}) + 1 = \left[\xi^{1h}_{gg}(r_{\rm co})+1\right] + \left[\xi^{2h}_{gg}(r_{\rm co})+1\right],
\end{equation}

\noindent where $1h$ and $2h$ refers to ``one-halo'' and ``two-halo''
terms, respectively. The one-halo correlation function is written as

\begin{displaymath}
1+\xi_{gg}^{1h}(r_{\rm co})=  \hspace{0.65\columnwidth}
 \end{displaymath}
\begin{equation}
\quad \frac{1}{2\pi r_{\rm co}^2\overline{n}_g^2}
              \int dM_h \frac{dn}{dM_h}\frac{\langle N(N-1)\rangle_M}{2}
              \frac{1}{2R_{h}} F^\prime\left(\frac{r_{\rm co}}{2R_{h}}\right),
\end{equation}

\noindent where $\bar{n}_g$ is the space density of galaxies in the
sample being modeled, $\langle N(N-1)\rangle_M$ is the second moment
of the distribution of galaxies within halos as a function of halo
mass, and $F^\prime(x)$ is the radial distribution of pairs within the
halo normalized to unity. Within a halo, pairs of galaxies can be
between the central galaxy and a satellite, or between two satellite
galaxies. The radial pair profile is different for these two
combinations, thus we express their relative contributions to
$\xi_{gg}^{1h}(r_{\rm co})$ as

\begin{eqnarray}
        \frac{\langle N(N-1)\rangle_M}{2}F^\prime(x) & = & \langle N_{\rm
        cen}N_{\rm sat} \rangle_MF^\prime_{\rm cs}(x) \nonumber \\ & & 
	+ \frac{\langle N_{\rm sat}(N_{\rm sat}-1)\rangle_M}{2}F^\prime_{\rm ss}(x),\nonumber \\
&&
\end{eqnarray}

\noindent where $F^\prime_{cs}$ is the pair distribution for
central-satellite pairs and $F^\prime_{ss}$ is the equivalent for
satellite-satellite pairs. The former is related to the density
profile of satellite galaxies, and the latter is related to the
density profile convolved with itself (analytic expressions for this
convolution can be found in \citealt{Sheth:2001a}). Here we assume
that the radial distribution of satellite galaxies is the same as the
dark matter, for which we assume the profile form of
Navarro-Frenk-White \citep[NFW;][]{Navarro:1997} using the
mass-concentration relation of \citet[][]{Munoz-Cuartas:2010}. Since
we assume that satellites trace the dark matter, $F^\prime_{cs}$ will
be equal to the quantity $F^\prime_{c}$ that we will introduce in
Equation \ref{eq:nf} and $F^\prime_{ss}$ will be equal to
$F^\prime_{s}$. 

Central galaxies only exist as one or zero objects in a halo, thus
they have no second moment. For the second moment of satellite
galaxies, we assume Poisson statistics about $\langle N_{\rm
  sat}\rangle$, which is in good agreement with results from numerical
simulations (\citealt{Kravtsov:2004,Zheng:2005}). Possible deviations
from Poisson behavior \citep[][]{Busha:2010,Boylan-Kolchin:2010}
mainly affect clustering for galaxy samples where a majority of the
satellites originate in halos with $M_h<M_{\rm sat}$ because in this
case, $\nsat$ drops to $\nsat\lesssim 1$. Luminous Red Galaxies (LRGs)
fall into this category for example. Indeed, satellite galaxies in LRG
samples mainly originate in $M_h<M_{\rm sat}$ halos because $M_{\rm
  sat}$ is close to the exponential cut-off in the halo mass function
(for LRGs, $M_{\rm sat} \sim 4\times10^{14}$ M$_{\odot}$). However,
for the types of samples that we consider in Paper II, deviations from
Poisson statistics should not significantly effect the clustering
predictions of the HOD.

A detailed description of the two-halo term can be found in
\cite{Tinker:2005}. Briefly, in the regime where $r>R_{h}$ of
massive halos, this term can be expressed as

\begin{equation}
\label{e.2halo_approx}
\xi_{gg}^{2h}(r_{\rm co}) = b_g^2\zeta^2(r_{\rm co})\xi_m(r_{\rm co}),
\end{equation}

\noindent where $\xi_m(r_{\rm co})$ is the non-linear matter correlation
function and $b_g$ is the large-scale bias of galaxies in the sample,
and $\zeta(r_{\rm co})$ is the scale dependence of dark matter halo bias. For
$\xi_m(r_{\rm co})$, we use the fitting function of \cite{Smith:2003}. For
$\zeta(r_{\rm co})$, we use the fitting function of \cite{Tinker:2005}. The
galaxy bias is computed from the HOD by

\begin{equation}
  b_g = \bar{n}_g^{-1}\int b(M_h)\langle N\rangle_M\frac{dn}{dM_h}dM_h,
\end{equation}

\noindent where $dn/dM_h$ is the halo mass function, for which we use
\citet{Tinker:2008}, and $b(M_h)$ is the halo bias function, for which
we use \cite{Tinker:2010}.

In the regime where $r< R_{h}$, Equation \ref{e.2halo_approx}
breaks down due to the effects of halo exclusion; i.e., the effect
that the center of one halo cannot exist within the virial radius of
another halo (and still be considered a ``two-halo'' pair). This is
explained in detail in \cite{Tinker:2005}. Because the mass function
and bias relation used in this analysis are taken from numerical
results based on spherical-overdensity (SO) halo catalogs
(\citealt{Tinker:2008, Tinker:2010}), the halo exclusion must be
modified to match this halo definition. In the SO halo finding
algorithm of \cite{Tinker:2008}, halos are allowed to overlap so long
as the center of one halo is not contained within the radius of
another halo. Thus, the minimum separation of two halos with radii
$R_1\ge R_2$ is $R_1$, rather than the sum of the two radii, as done
in \cite{Tinker:2005}. For a projected statistic like $w(\theta)$,
this makes only a small difference in clustering at the 1-halo 2-halo
transition, but significantly speeds up computation of the 2-halo
term.

Once we have calculated $\xi_{gg}(r_{\rm co})$ for a given HOD model,
we compute the observable $w(\theta)$ by:

\begin{equation}
\label{e.wth}
w(\theta) = \int dz\,N^2(z)\, \frac{dr_{\rm co}}{dz} \int dx\,
\xi\left(\sqrt{x^2 + r_{\rm co}^2\theta^2}\right),
\end{equation}

\noindent where $N(z)$ is the normalized redshift distribution of the
galaxy sample, $r_{\rm co}$ is the comoving radial coordinate at redshift $z$
and $dr_{\rm co}/dz = (c/H_0)/\sqrt{\Omega_{\rm
    m}(1+z)^3+\Omega_\Lambda}$.

\begin{figure*}[htb]
\epsscale{1.1}
\plotone{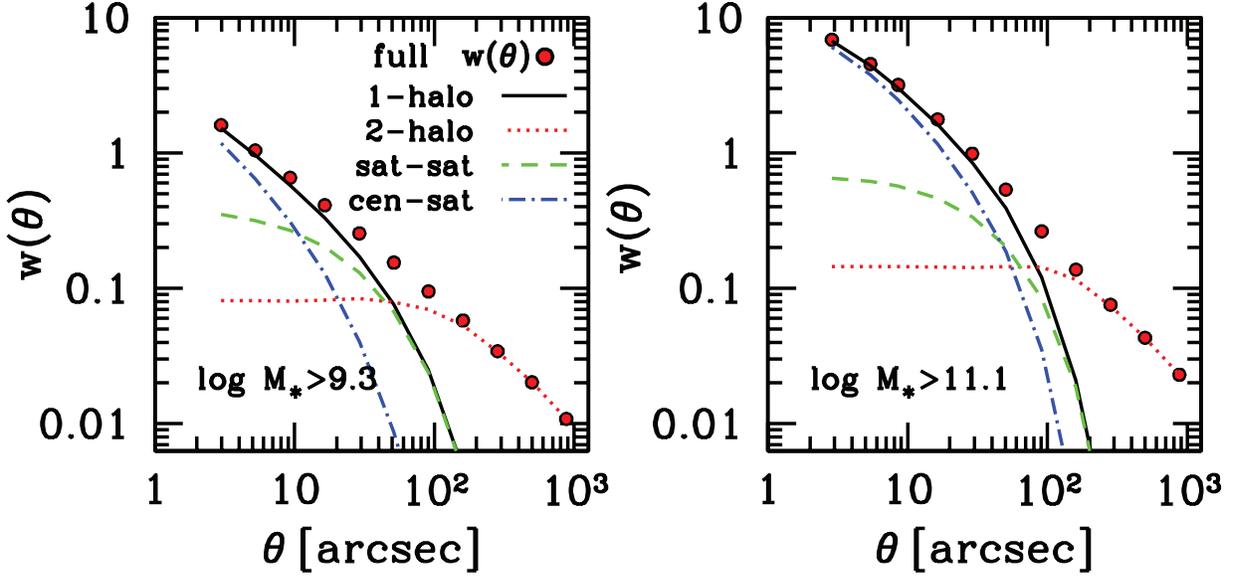}
\caption{The angular correlation function for two different stellar
  mass thresholds. The filled points are the full HOD calculation for
  $w(\theta)$. The different curves break the calculation into its
  distinct parts: The solid curve is the 1-halo term, the dotted curve
  is the two halo term. We further break the 1-halo term into the
  relative contribution of central-satellite galaxy pairs ( dash-dot
  curve) and satellite-satellite galaxy pairs (long-dash curve). For
  more massive galaxy samples, the one-halo term is more prominent and
  it is dominated by central-satellite pair counts.}
\label{illustration_wtheta1}
\end{figure*}

Figure \ref{illustration_wtheta1} shows the breakdown of the angular
correlation function into the one-halo and two-halo terms for low-mass
and high-mass galaxy samples.

\subsection{Analytical modeling of $\xi_{gm}$ and $\Delta\Sigma$}\label{model_for_xi_gm}

We have shown in $\S$ \ref{ds_correlation_funct} that the lensing
observable, $\Delta\Sigma$, can be obtained from $\xi_{gm}$ by
performing two integrals (Equations \ref{dsigma1_bis} and
\ref{dsigma2}). Following the approach of \citet[][]{Yoo:2006},
$\xi_{gm}$ is computed from the HOD and $\Delta\Sigma$ is obtained by
combining Equations \ref{dsigma}, \ref{dsigma1_bis} and
\ref{dsigma2}. Thus $\Delta\Sigma$ is fully specified given our
model. Note that the calculation of $\xi_{gm}$ is performed in
comoving units and then the projections of Equations \ref{dsigma1_bis}
and \ref{dsigma2} to obtain $\Delta\Sigma$ are performed in physical
units (to match our measured g-g lensing signal which is computed in
physical units in Paper II).

Typically, $\xi_{gm}$ is decomposed into a one-halo and a two-halo
term

\begin{equation}
1+\xi_{gm}(r_{\rm co}) = [1+\xi_{gm}^{1h}(r_{\rm co})]+[1+\xi_{gm}^{2h}(r_{\rm co})],
\end{equation}

\noindent where the one-halo term represents galaxy-matter pairs from
single halos and is dominant on small scales ($\lesssim$Mpc) and the
two-halo term corresponds to pairs from distinct halos and is dominant
on larger scales ($\gtrsim$Mpc). The one-halo term is obtained according
to:

\begin{displaymath}
 1+\xi_{gm}^{1h}(r_{\rm co}) =  \hspace{0.65\columnwidth}
 \end{displaymath}
\begin{equation}
 \label{eq:1h}
\quad  \frac{1}{4\pi r_{\rm co}^2 \overline{n}_g}\int_{0}^{\infty}\frac{{\rm
    d}n}{{\rm
    d}M_h}\frac{M_h}{\overline{\rho}}\frac{1}{2R_{h}}\langle
N_{tot}\rangle F'\left(\frac{r_{\rm co}}{2R_{h}}\right){\rm d}M_h .
\end{equation}

Further details about the origin of Equation \ref{eq:1h} are presented
in the Appendix.

The product $\overline{n}_g F'$ is split into two terms:

\begin{equation}
\overline{n}_g F'(x)=\overline{n}_c F'_{c}(x)+ \overline{n}_s F'_{s}(x),
\label{eq:nf}
\end{equation}

\noindent where $F'_{c}$ is linked to the density profiles of dark
matter halos (see Appendix) and $F'_{s}$ is related to the convolution
of the dark matter density profile and the satellite galaxy
distribution. To calculate $F'_{c}$ we assume spherical NFW profiles
truncated at $R_{h}$ and we adopt the \citet{Munoz-Cuartas:2010}
mass-concentration relation for a WMAP5 cosmology. To calculate
$F'_{s}$ we assume that the satellite galaxy distribution follows the
dark matter distribution. Given this assumption, $F'_{s}$ is simply
related to the convolution of the NFW profile with itself.

In this paper, we neglect the contribution to $\Delta\Sigma$ from
sub-halos; \citet[][]{Yoo:2006} have shown this component to be
negligible at the 10\% level.

We calculate the two-halo term as described in \cite{Yoo:2006}, with
two major exceptions. First, as stated in the previous section, we are
using the halo exclusion from the SO halo definition. Second, because
the halo mass function of \cite{Tinker:2008} and halo bias function of
\cite{Tinker:2010} are normalized such that integrating over all $M_h$
produces the mean matter density and a bias of unity, there is no need
to employ the ``break mass'' from \citet{Yoo:2006} (see their Equation
16). In the limit where $r>R_{h}$ of massive halos,

\begin{equation}
\xi_{gm}^{2h}(r_{\rm co}) = b_g\zeta(r_{\rm co})\xi_m(r_{\rm co}),
\end{equation}

\noindent analogous to Equation \ref{e.2halo_approx}.

In addition to the two terms presented above, we add another component
to the modeling of $\Delta\Sigma$ which is absent in
\citet[][]{Yoo:2006}, namely the contribution to $\Delta\Sigma$ from
the baryons of the central galaxy which can be non negligible on very
small scales ($<$ 50 kpc). Although the baryons typically follow
S\'{e}rsic profiles \citep{Sersic:1963}, at the scales of interest for
this study, well above a few effective radii ($>$ 20 kpc), the lensing
contribution of the baryons can be modeled by a simple point-source,
scaled to $\langle M_{*}\rangle$, the average stellar mass of the
galaxies in the sample:

\begin{equation}
  \Delta\Sigma_{\rm stellar}(r)=\frac{\langle M_{*}\rangle }{\pi r^2}.
\end{equation}

In total, the final g-g lensing signal is modelled as the sum of three
terms: $\Delta\Sigma_{\rm tot}=\Delta\Sigma_{\rm
  stellar}+\Delta\Sigma_{1h}+\Delta\Sigma_{2h}$. Note that the one
halo and two halo terms can also be decomposed into central and
satellite contributions but for simplicity, we have grouped these
terms together.

\begin{figure*}[htb]
\epsscale{1.1} 
\plotone{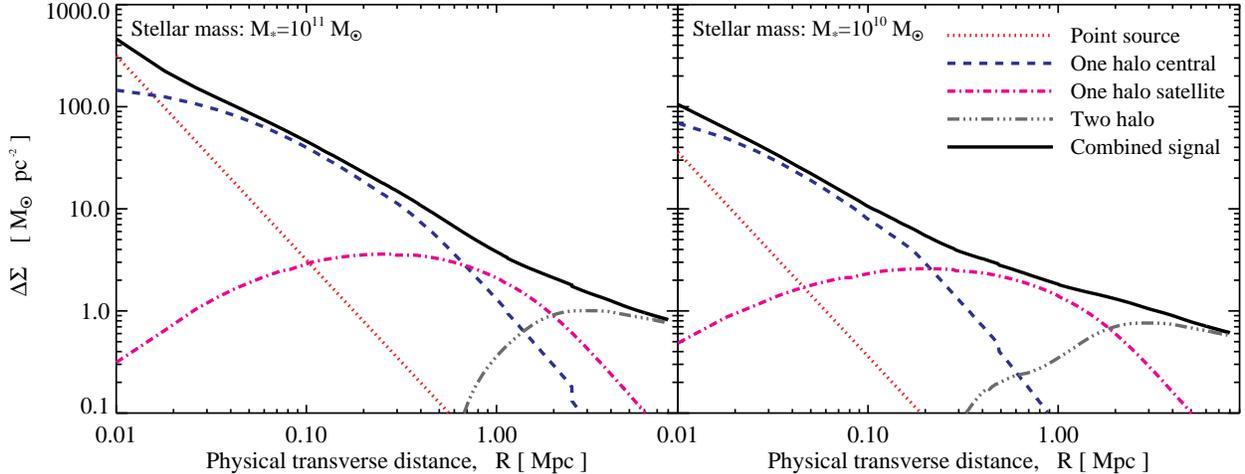}
\caption{Illustration of the various terms that contribute to a g-g
  lensing signal. The solid black curve shows the total g-g lensing
  signal which can be decomposed into a sum of terms that contribute
  at various scales. On small scales ($\sim 10$ kpc), the signal is
  dominated by the baryonic content of galaxies represented by the red
  curve (dot-dot). At intermediate radii ($\sim 200$ kpc), dark matter
  halos come into play as shown by the blue (dash-dash) and magenta
  (dash-dot) curve. The former represents a NFW profile and the
  later is due to a contribution from satellite galaxies. On large
  scales ($>3 $ Mpc), the g-g lensing follows the dark matter linear
  auto-correlation function scaled by the bias factor as depicted by
  the grey curve (dash-dot-dot-dot). It is interesting to note that the
  total g-g lensing signal is roughly a power law, despite the fact
  that the various contributing components deviate strongly from power
  laws.}
\label{illustration_gg_lensing}
\end{figure*}

In order to illustrate the various terms that contribute to the g-g
lensing signal, we have plotted the signal in Figure
\ref{illustration_gg_lensing}.

\section{Influence of the model parameters on the observables}\label{parameter_influance}

\begin{figure*}[p]
\epsscale{1.15}
\plotone{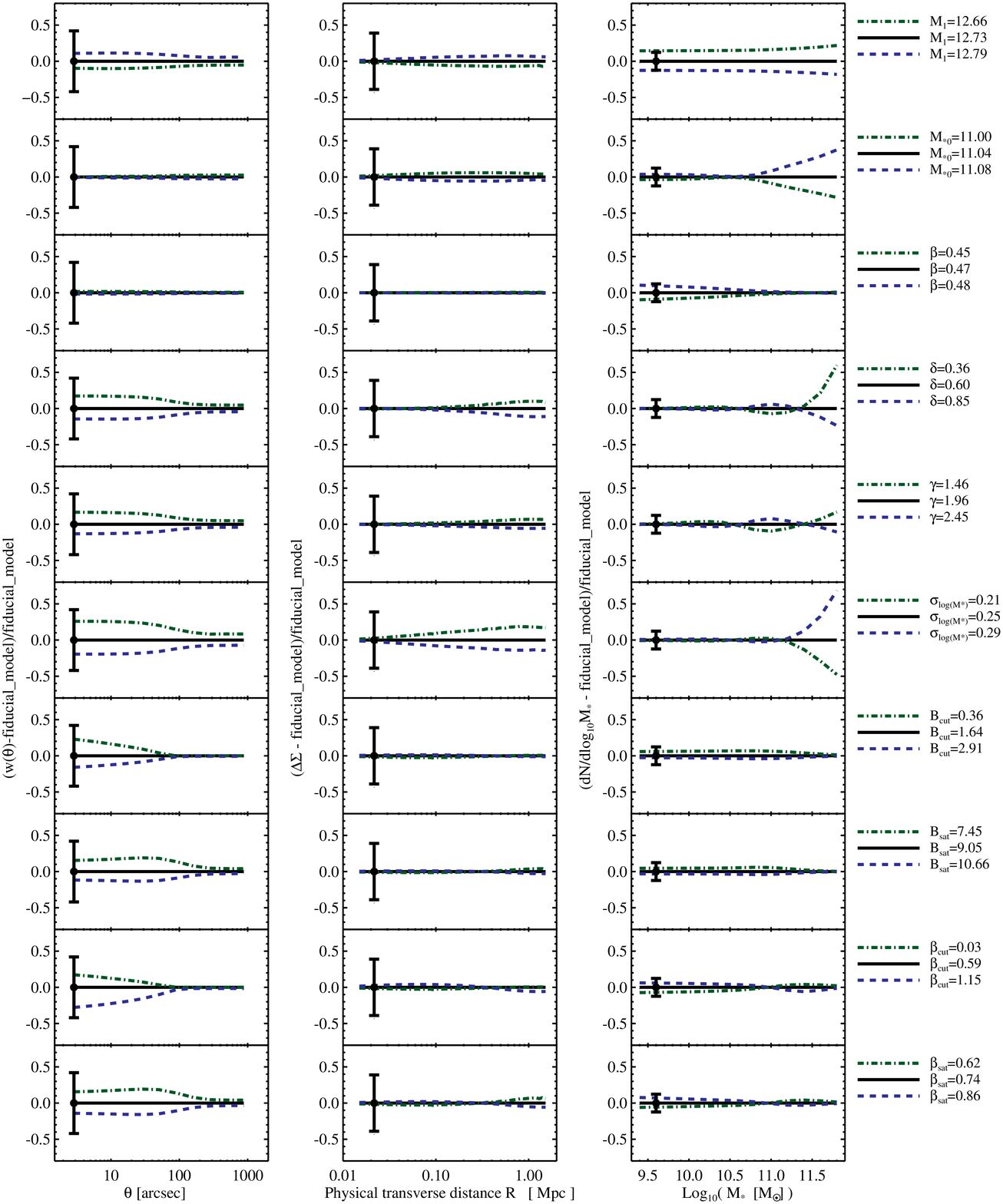}
\caption{Effect of varying each of the 10 parameters in turn by
  $2\sigma$ around the best fit model from Paper II for $0.48<z<0.74$
  where $\sigma$ is the fitted error on the parameter in question. In
  the left hand column of figures, we show how the predicted
  $w(\theta)$ signal varies for $\log_{10}(M_*)>11.1$ (a high stellar
  mass threshold). In the middle column of figures, we show how the
  predicted $\Delta\Sigma$ signal varies for $11.29<\log_{10}(M_*)<12$
  (a high stellar mass bin) and the right hand column shows the
  predicted variations for SMF down to $\log_{10}(M_*)=9.3$. To
  highlight the effects of the parameter variation, in all three cases
  we have plotted the model minus the fiducial best-fit model divided
  by the fiducial best-fit model. The data point with an error bar
  represents the typical error bar for each observable for a
  COSMOS-like survey.}
\label{p_variation_massive}
\end{figure*}

\begin{figure*}[p]
\epsscale{1.15}
\plotone{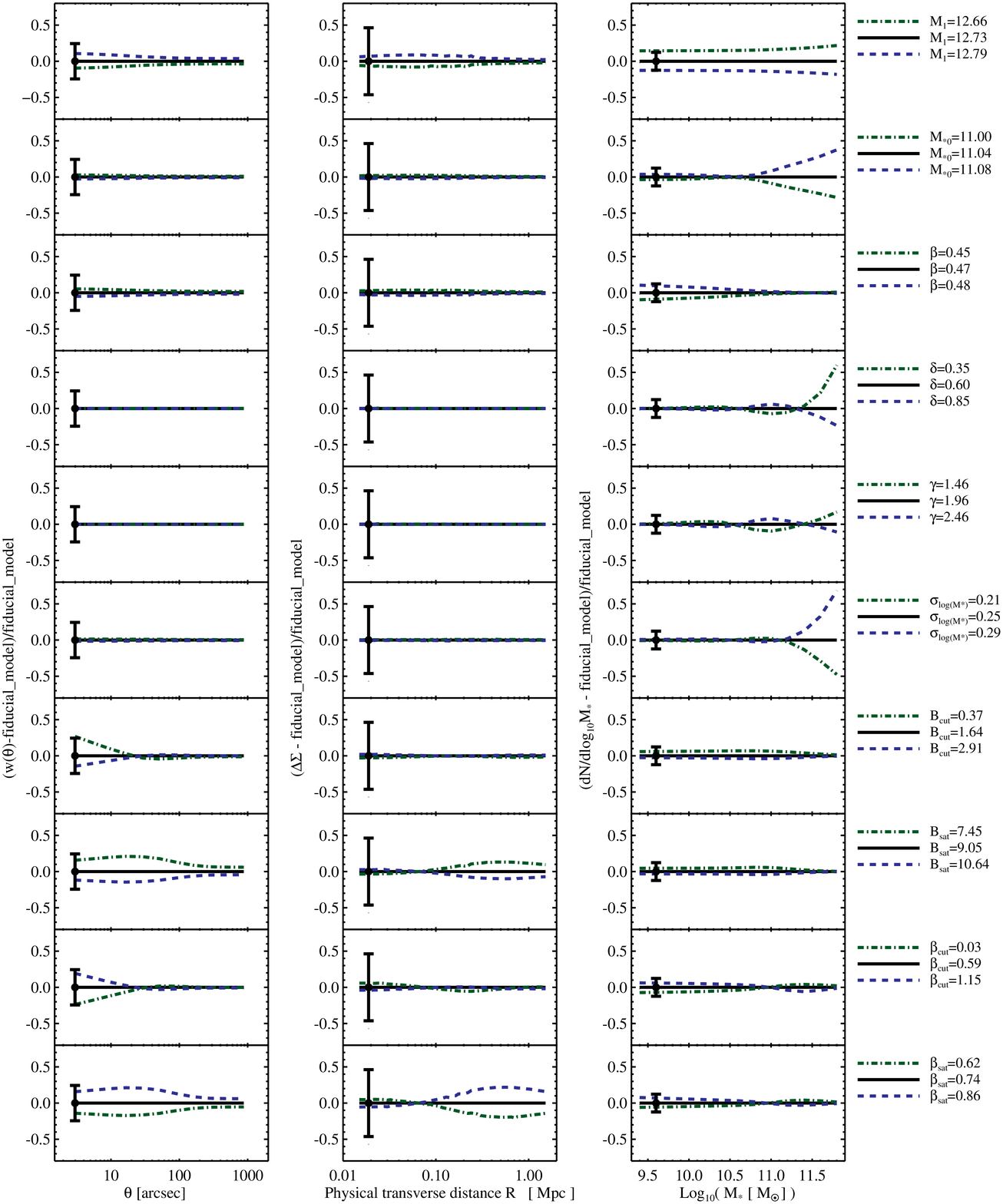}
\caption{Same as Figure \ref{p_variation_massive} but the left hand
  column of figures now shows $w(\theta)$ for $\log_{10}(M_*)> 9.3$ (a
  low stellar mass threshold) and the middle column of figures now
  shows $\Delta\Sigma$ for $9.8<\log_{10}(M_*)<10.3$ (a low stellar
  mass bin). The right hand column (depicting the SMF) is identical to
  Figure \ref{p_variation_massive}.}
\label{p_variation_low}
\end{figure*}

In the previous section we have outlined how our model can be used to
analytically predict the SMF, g-g lensing, and clustering signals. We
will now investigate how each parameter in the model affects the three
observables. For this exercise, we adopt the best fit model parameters
for $0.48<z<0.74$ from Paper II and we vary each parameter in turn by
$2\sigma$ around the best fit model. For this section, we assume that
$\sigma_{\rm log M_{*}}$ is constant. We also assume that $\alpha_{\rm
  sat}$ is constant and we set $\alpha_{\rm sat}=1$ in this section
since this is also the assumption that we make in Paper II. In total,
we therefore study the effects of ten parameters: $M_{1}$, $M_{*,0}$,
$\beta$, $\delta$, $\gamma$, $\sigma_{\rm log M_{*}}$, $\beta_{\rm
  sat}$, $B_{\rm sat}$, $\beta_{\rm cut}$, and $B_{\rm cut}$. The
results are shown in Figures \ref{p_variation_massive} and
\ref{p_variation_low} and are described in further detail below.


\subsection{Effect of parameters on the SMF}

The influence of each model parameter on the observed SMF is shown in
the right hand column in Figure \ref{p_variation_massive} (the same
column is reproduced in Figure \ref{p_variation_low}). The data point
with an error bar represents the typical error bar for a COSMOS-like
survey where the error bar includes sample variance computed from a
series of mock catalogs (described in the following section). The
first point worth mentioning here is that the errors on the SMF are
relatively small compared to the clustering and the lensing. It is
always the case that a measurement of a one-point statistic from a
given set of data is more precise than a measurement of a two-point
(or higher) statistic. The implies that the SMF will in general play
an important role in constraining the parameters of the SHMR. However,
in follow-up work we will investigate the sensitivity of this probe
combination to cosmological parameters and models of modified gravity
for example. In these studies, the clustering and the g-g lensing will
play a critical role despite their typically larger errors bars.

A second noteworthy point in Figure \ref{p_variation_massive} is that
the SMF appears to be sensitive to all ten parameters whereas certain
parameters such as $M_{*,0}$ and $\beta$ have very little effect on the
clustering and lensing signals. Coupled with the fact that the SMF has
fairly small error bars, this implies that the SMF will have quite a
lot of constraining power on the overall model compared to the g-g
lensing for example.

The effects of $M_{1}$ and $M_{*,0}$ on the SMF are fairly
intuitive. $M_{1}$ roughly induces an up/down shift in the amplitude
of the SMF and $M_{*,0}$ corresponds to a left/right shift in the
SMF. A larger value of $B_{\rm sat}$ implies that there are fewer
satellite galaxies in high mass halos for a given stellar mass
threshold. Thus we see that an increase in $B_{\rm sat}$ corresponds
to a decrease in the amplitude of the SMF due to the fact that the
contribution from satellite galaxies has decreased. Similar arguments
apply to $B_{\rm cut}$. From Figure \ref{p_variation_massive} we can
anticipate that if the SMF were only used to constrain this model,
degeneracies would occur between $B_{\rm sat}$, $B_{\rm cut}$, and
$M_{1}$. Fortunately, the satellite parameters have a significant
effect on both the clustering and the g-g lensing so this degeneracy
should be broken when all three probes are used in conjunction.

\begin{figure*}[htb]
\epsscale{1.0}
\plotone{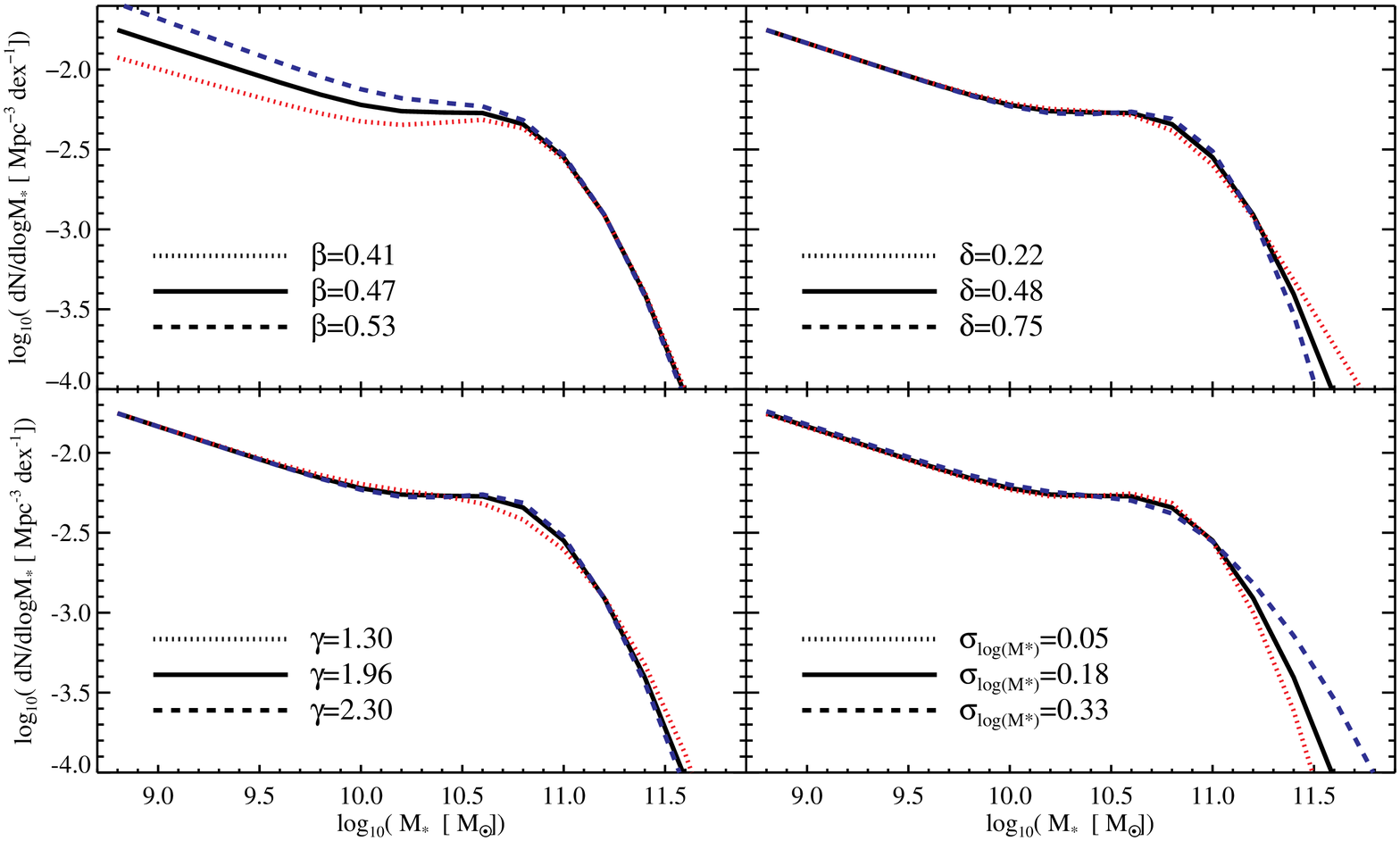}
\caption{Effects of $\beta$, $\delta$, $\gamma$, and $\sigma_{\rm log
    M_{*}}$ on the shape of the observed SMF. Left upper panel:
  $\beta$ determines the low mass slope of the SMF. Right upper panel:
  $\delta$ affects the knee and the high-mass slope of the SMF. Left
  lower panel: $\gamma$ affects the knee of the SMF but $\gamma$ also
  affects the ``plateau'' feature that has been observed in the SMF at
  $10<\log_{10}(M_*)<10.5$ (see discussion and references in
  \citealt[][]{Drory:2009}). Right lower panel: $\sigma_{\rm log M_{*}}$
  affects the high mass slope of the SMF but also affects the
  ``plateau'' feature. A larger value of $\sigma_{\rm log M_{*}}$
  leads to an inflated observed SMF at the high mass end. This effect
  is also commonly referred to as Eddington bias.}
\label{smf_dip_feature}
\end{figure*}

In Figure \ref{smf_dip_feature} we highlight the effects of four
particular parameters on the SMF: $\beta$, $\gamma$, $\delta$, and
$\sigma_{\rm log M_{*}}$. The $\beta$ parameter affects the low-mass
slope of the SMF so that a larger value of $\beta$ corresponds to a
steeper low-mass slope in the SMF. The $\gamma$ parameter (we recall
that this controls the transition region of the SHMR as can be seen
from Figure \ref{mh_ms_vary_p}) has an interesting effect since it
regulates a ``plateau'' feature in the SMF at $\log_{10}(M_*)\sim
10.5$. In fact, this feature in the SMF has been noticed already and
discussed in detail for example in \citet[][]{Drory:2009}. In
\citet[][]{Drory:2009}, this feature was described as a ``dip''
because at these scales, the SMF is below the best fit Schecter
function. However, we note that ``dip'' is a somewhat misleading name
for this feature since it could also be taken to mean that ${\rm
  d}N/{\rm d}\log_{10}(M_*)$ does not decrease monotonically with
$M_*$. A close inspection of our SMFs in Paper II shows that there is
no evidence in the data for an actual ``dip'' in ${\rm d}N/{\rm
  d}\log_{10}(M_*)$. Instead, the data are more consistent with a
flattening of ${\rm d}N/{\rm d}\log_{10}(M_*)$ around
$\log_{10}(M_*)\sim 10.5$. Therefore, we would like to suggest that
this feature should be described as a ``plateau'' in the SMF rather
than a ``dip''.

\begin{figure*}[htb]
\epsscale{1.09} 
\plotone{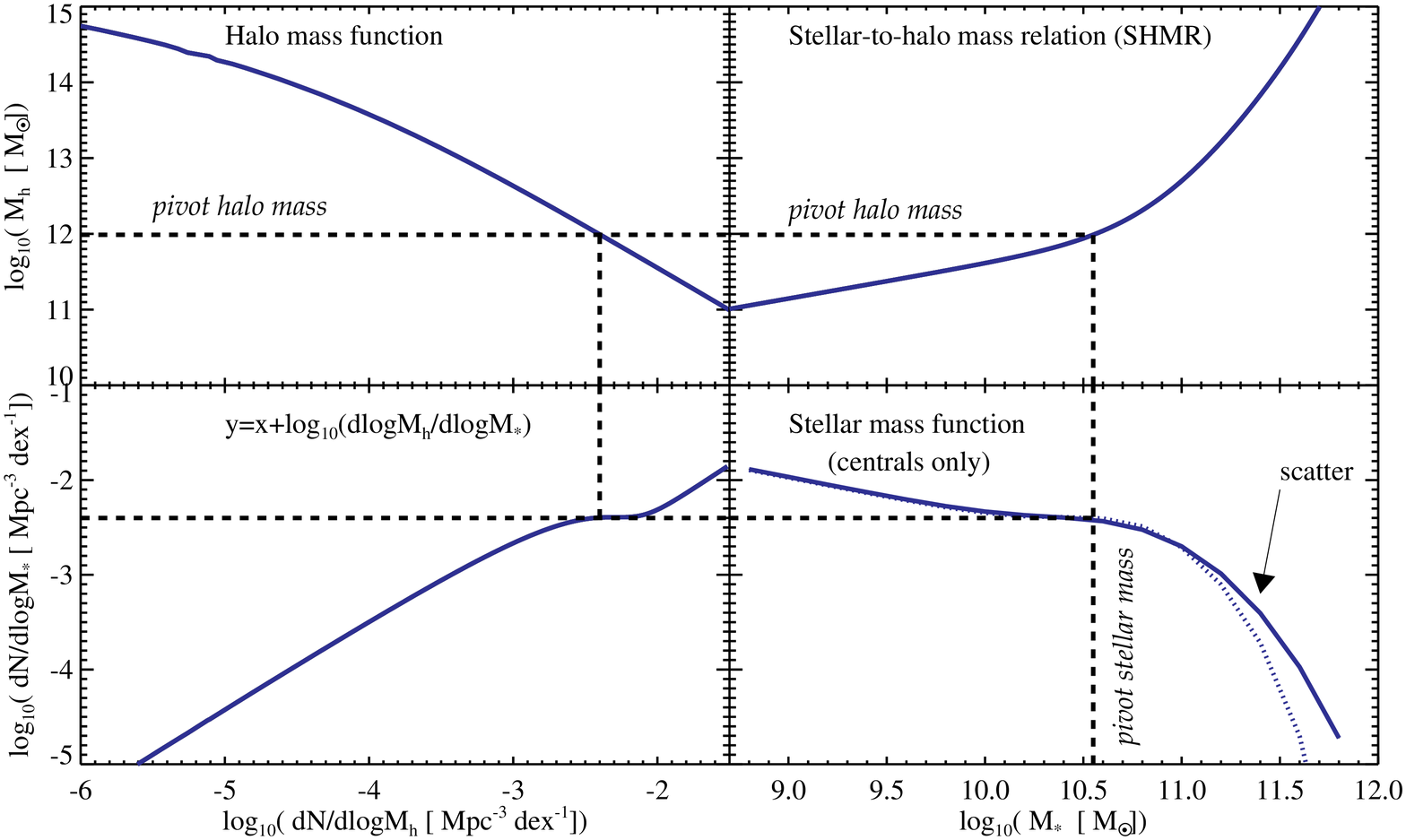}
\caption{Illustration of the link between the halo mass function
  (upper left panel), the SHMR (upper right panel) and the SMF (lower
  right panel). The contribution to the SMF from satellite galaxies
  has been subtracted so that the SMF in the lower right hand panel
  only depicts the contribution from central galaxies. The dot-dot
  line shows the SMF when $\sigma_{\rm log M_{*}}=0$. This figure is
  designed that all three quantities can be linked by drawing a
  ``box'' between the four panels as shown here by example with the
  dash-dash lines at the location of the pivot stellar mass. We note
  that this will only work when the SMF with zero scatter (dot-dot
  line) is considered. This figure shows for example that the location
  of the pivot stellar mass is coincident with the location of the
  ``plateau'' feature in the SMF.}
\label{smf_dip_feature2}
\end{figure*}

In Figure \ref{smf_dip_feature2} we show the link between the dark
matter halo mass function, the SHMR, and the SMF. In this Figure, we
have used the fact that ${\rm d}N/{\rm d}\log_{10}M_* = {\rm d}N/{\rm
  d}\log_{10}M_h\times({\rm d}\log_{10}M_h/{\rm d}\log_{10}M_*)$ so
that the various functions can be linked ``by eye'' by drawing a box
between the four different panels. We have illustrated how to link the
various functions with the dash-dash lines in Figure
\ref{smf_dip_feature2} at the scale of the pivot stellar mass. Figure
\ref{smf_dip_feature2} shows that the ``plateau'' feature is caused by
the transition that occurs in the SHMR at $M_h\sim 10^{12}$
M$_{\odot}$ from a low-mass power-law regime to a sub-exponential
function at higher stellar mass.

Finally, the scatter in stellar mass at fixed halo mass has a
noticeable effect on the SMF at the high mass end which is also
commonly referred to as Eddington bias. A larger value of $\sigma_{\rm
  log M_{*}}$ will lead to an inflated observed SMF at large stellar
masses.

\subsection{Effect of parameters on g-g lensing}

The g-g lensing signal is dominated by the central one-halo term
roughly on scales below $0.3$ Mpc and by the satellite one-halo term
roughly on scales above $0.3$ Mpc and below a few Mpc (see Figure
\ref{illustration_gg_lensing}). Thus, the effects of the four
parameters that regulate the satellite occupation function ($B_{\rm
  sat}$, $B_{\rm cut}$, $\beta_{\rm sat}$, $\beta_{\rm cut}$) have a
strong scale-dependant effect on the g-g lensing signal. For example,
$B_{\rm sat}$ controls the power-law amplitude of $\nsat$. A smaller
value of $B_{\rm sat}$ will reduce the ratio $M_{\rm sat}/\fshmr$ and
will therefore increase the number of satellites in a sample. This
will lead to an increase in the g-g lensing signal from 0.3-1 Mpc due
to the increased amplitude of the one-halo satellite term. Similarly,
a smaller value of $\beta_{\rm sat}$ will also reduce the ratio
$M_{\rm sat}/\fshmr$ and by consequence, will increase the one-halo
satellite term. This effect is more pronounced in Figure
\ref{p_variation_low} which illustrates a low-stellar mass sample
compared to Figure \ref{p_variation_massive} which illustrates a high
stellar mass sample.

Another parameter worth discussing here is $\sigma_{\rm log
  M_{*}}$. Figure \ref{p_variation_massive} demonstrates that
$\sigma_{\rm log M_{*}}$ has a stronger effect on the lensing signal
for high stellar mass samples compared to low stellar mass samples. As
discussed in $\S$ \ref{scatter_shmr}, this is simply due to the fact
that the data are binned according to $M_*$. The observables are
therefore sensitive to the scatter in halo mass at fixed stellar mass,
$\sigma_{\rm log M_h}$. At fixed $\sigma_{\rm log M_{*}}$,
$\sigma_{\rm log M_h}$ will increase with $M_*$. As a result, the
effects of scatter are more prominent in g-g lensing measurements for
high stellar mass samples.

\subsection{Effect of parameters on clustering}

To first order, the stellar mass function and the clustering of
galaxies are tethered; more massive halos are both more clustered and
less abundant. This is also true of galaxies because rare, massive
galaxies live in such halos. If the amplitude of the stellar mass
function increases, the clustering as a function of stellar mass
decreases. This is especially true at masses above the knee in the
stellar mass function. From Figure \ref{smf_dip_feature}, increasing
$\delta$ or $\sigma_{\rm log M_{*}}$ increases the abundance of
high-mass galaxies. Given that the number of halos is fixed, this can
only mean that massive galaxies are occupying less massive, less
clustered halos.

There are several parameters that have a direct influence on the
clustering of galaxies without changing the stellar mass function
appreciably. The parameters $B_{\rm sat}$ and $\beta_{\rm sat}$ are
the most important in this regard. They control the ``shoulder'' in
the HOD, defined conceptually as the increase in halo mass, relative
to $\fshmr$, before satellites begin to enter the
sample. Quantitatively, this is expressed as the ratio $M_{\rm sat}
/\fshmr$, as shown in Equation \ref{e.nsat}. Reducing this ratio
increases the number of satellite galaxies in a sample, which in turn
increases the large-scale bias of a sample and significantly enhances
the clustering within the one-halo term. The parameters $B_{\rm cut}$
and $\beta_{\rm cut}$ have a more subtle effect on clustering. If the
cutoff mass, defined by Equation \ref{mcut_eq}, is below $\fshmr$,
then $M_{\rm cut}$ has no effect on clustering. But as $M_{\rm cut}$
increases, satellite galaxies are removed from low-mass halos. If the
density of satellites is held fixed, increasing $M_{\rm cut}$
redistributes satellites into more massive halos. This will increase
the large-scale bias and change the shape of the one-halo term such
that the correlation function deviates from a pure power law form (see
the Appendix in \citealt{Zheng:2009}).


\section{Mock catalogs, sample variance, and covariance}\label{mocks}

In this section we construct mock catalogs in order to investigate the
effects of sample variance and covariance associated with measurements
of g-g lensing, clustering, and the SMF. Sample variance occurs due to
the finite nature of the volume encompassed by any given
survey. Because of limited volume, any given survey may yield a biased
measurement of the number density of galaxies and halos compared to
the full universe. The error bars on all three observables must
therefore reflect this additional source of error. Also, the
data-points in all three observables will be correlated to some
degree. Consider the SMF for example. A region of space with high
matter density will have an increased abundance of galaxies of nearly
all masses. For $w(\theta)$, because it is a projection of
$\xi_{gg}(r)$---which is itself a correlated quantity---multiple
physical scales will contribute to each bin in $\theta$. For
$\Delta\Sigma$, we will show that the data points are correlated on
scales where satellite galaxies contribute to the lensing signal.

To investigate both the sample variance and the covariance associated
with all three observables, we use numerical simulations to construct
a series of mock catalogs for a COSMOS-like survey. Since the volume
of COSMOS is relatively small, the effects of variance and covariance
will be quite apparent (whereas the effects would decrease if we
simulated a larger fiducial survey) and so COSMOS is well suited for
our purpose. In addition, we will also use these mock catalogs in
Paper II to analyze the actual COSMOS data.

COSMOS-like mocks are created from a single simulation (named
``Consuelo'') 420 $h^{-1}$ Mpc on a side, resolved with 1400$^3$
particles, and a particle mass of 1.87$\times 10^{9}$ $h^{-1}$
M$_{\odot}$\footnotemark[3]\footnotetext[3]{In this paragraph, numbers
  are quoted for $H_0=100$ h km~s$^{-1}$~Mpc$^{-1}$}. This simulation
can robustly resolve halos with masses above $\sim 10^{11}$ $h^{-1}$
M$_{\odot}$ and is part of the Las Damas
suite\footnotemark[4]\footnotetext[4]{Details regarding this
  simulation can be found at {\tt
    http://lss.phy.vanderbilt.edu/lasdamas/simulations.html}} (McBride
et al. in prep). We create mocks for three redshift intervals:
$z_1=[0.22,0.48]$, $z_2=[0.48,0.74]$, and $z_3=[0.74,1]$. For each
redshift interval, we construct a series of mocks created from random
lines of sight through the simulation volume that have the same area
as COSMOS and the same comoving length for the given redshift
slice. This yields 405 independent mocks for the $z_1$ bin, 172 mocks
for the $z_2$ bin, and 109 mocks for the $z_3$ bin. For each redshift
bin, mocks are created from the simulation output at the median
redshift of the bin.

Halos within the simulation are identified with the friends-of-friends
halo finder (\citealt{Davis:1985}) with a linking length of b=0.2. For
each redshift interval, halos are populated with our best-fit model
from Paper II. We use the mock-to-mock variance and covariance to
estimate a covariance matrix for the stellar mass function, for
$w(\theta)$ (using a series of stellar mass thresholds), and for
$\Delta\Sigma$ (using a series of stellar mass bins). Although the
data points {\em between} the different quantities will be correlated
to some degree (as well as the bins in $w(\theta)$ and $\Delta\Sigma$)
we ignore that covariance as we do not have enough simulation volume
to estimate the {\it uber}-covariance matrix of all [N] data points in
each redshift bin.

\begin{figure*}[htb]
\epsscale{1.25}
\plotone{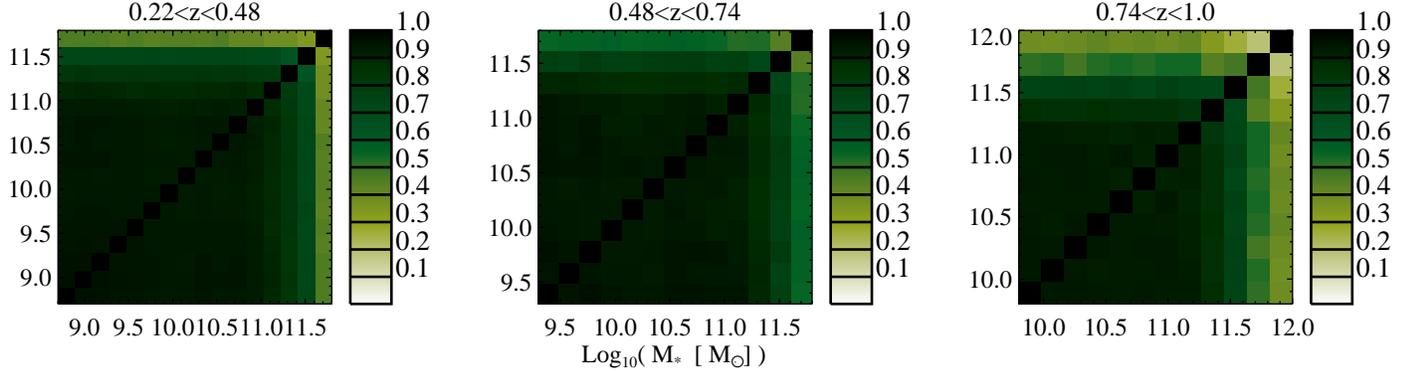}
\caption{Correlation coefficient matrix for the SMF in three redshift
  bins for a COSMOS-like survey. The effect of sample variance on the
  SMF is to correlate all of the data points so that globally, the SMF
  will shift up and down for different realizations of a COSMOS-like
  survey. }
\label{smf_covar_matrix}
\end{figure*}

Figure \ref{smf_covar_matrix} shows the correlation coefficient matrix
for the SMF in three redshifts bins for a COSMOS-like survey. The
first-order effect of sample variance on the SMF is to correlate all
of the data points so that globally, the SMF will shift up and down
for different realizations of a COSMOS-like survey.

\begin{figure*}[htb]
\epsscale{1.25} 
\plotone{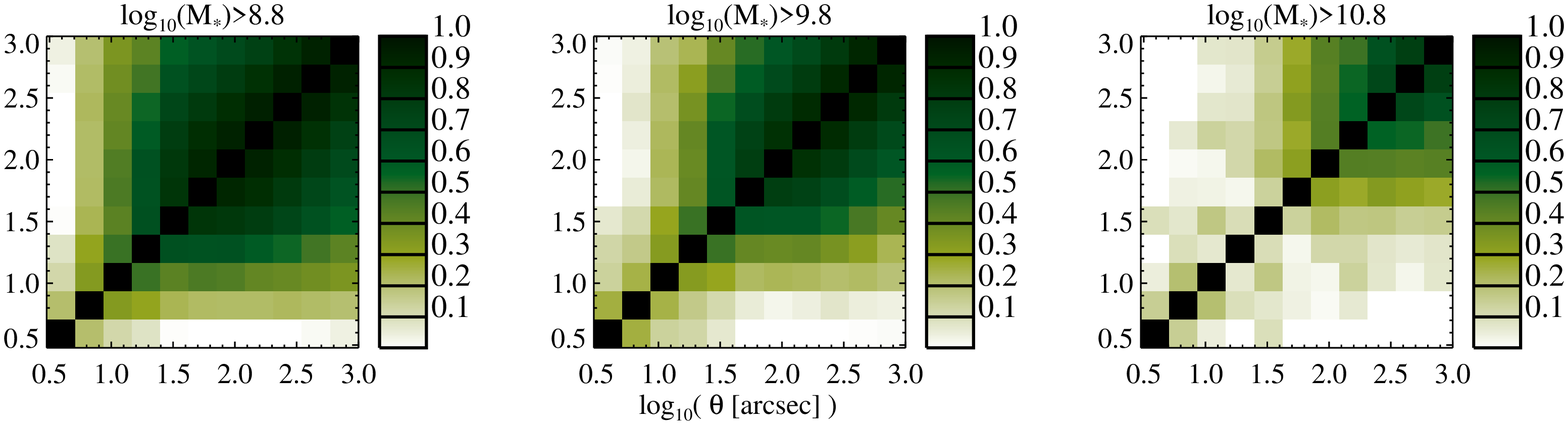}
\caption{Correlation coefficient matrix for galaxy clustering for
  $0.22<z<0.48$ and for several stellar mass thresholds.}
\label{clustering_covar_matrix}
\end{figure*}

Figure \ref{clustering_covar_matrix} shows the correlation coefficient
matrix for the galaxy clustering for $0.22<z<0.48$ and for three
stellar mass thresholds: $\log_{10}(M_*)>9.3$, $\log_{10}(M_*)>10.3$,
and $\log_{10}(M_*)>11.1$. The data are more correlated at larger
scales where galaxy pairs come from the two-halo term. As shown
earlier, clustering at these scales is proportional to the matter
clustering $\xi_m(r)$. Patches of the universe that exist in an over-
or under-density tend to have higher or lower clustering in their
matter. This will be reflected in the clustering of the halos and thus
the two-halo term for the galaxies. In the one-halo term, Poisson
fluctuations of the number of satellites become more important and the
data are less correlated at these scales. Overall, as the density of
the galaxy sample becomes smaller, shot noise will dominate on all
scales. This can be seen in the progression from left to right in the
examples in Figure \ref{clustering_covar_matrix}.

\begin{figure*}[htb]
\epsscale{1.25}
\plotone{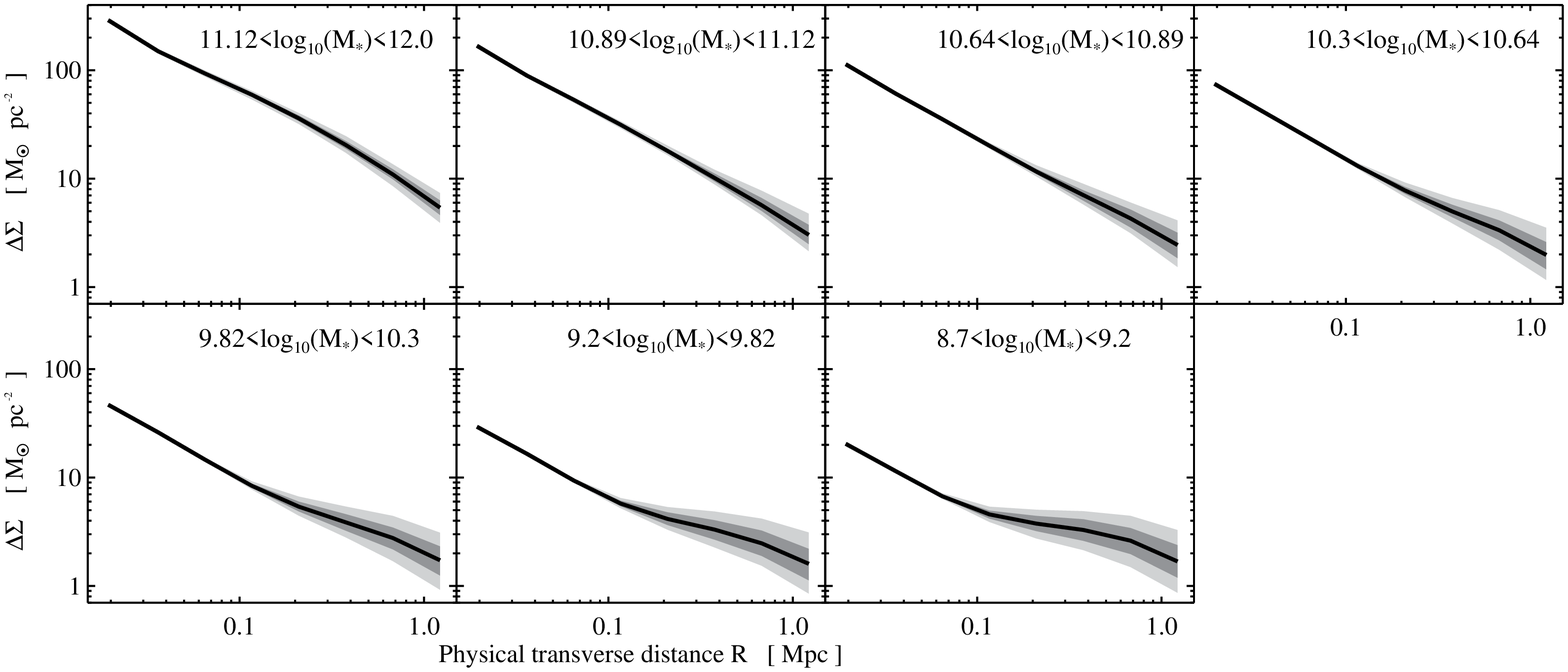}
\caption{Illustration of the effects of sample variance on the g-g
  lensing signals for a COSMOS-like survey. The light grey region
  represent the 2$\sigma$ variation between mocks and the dark grey
  region represents the 1$\sigma$ variation between mocks. Sample
  variance mainly affects the one-halo satellite term of the g-g
  lensing signal. For example, the large variance that can be seen in
  the bottom right panel ($8.7<\log_{10}(M*)<9.2$) from 100 kpc to 1
  Mpc is due to the one-halo satellite term. This can be explained by
  the fact that, for a given stellar mass sample, the parent halos of
  satellite galaxies are more rare at fixed volume than the halos of
  central galaxies.}
\label{gg_sample_variance}
\end{figure*}

\begin{figure*}[htb]
\epsscale{1.2}
\plotone{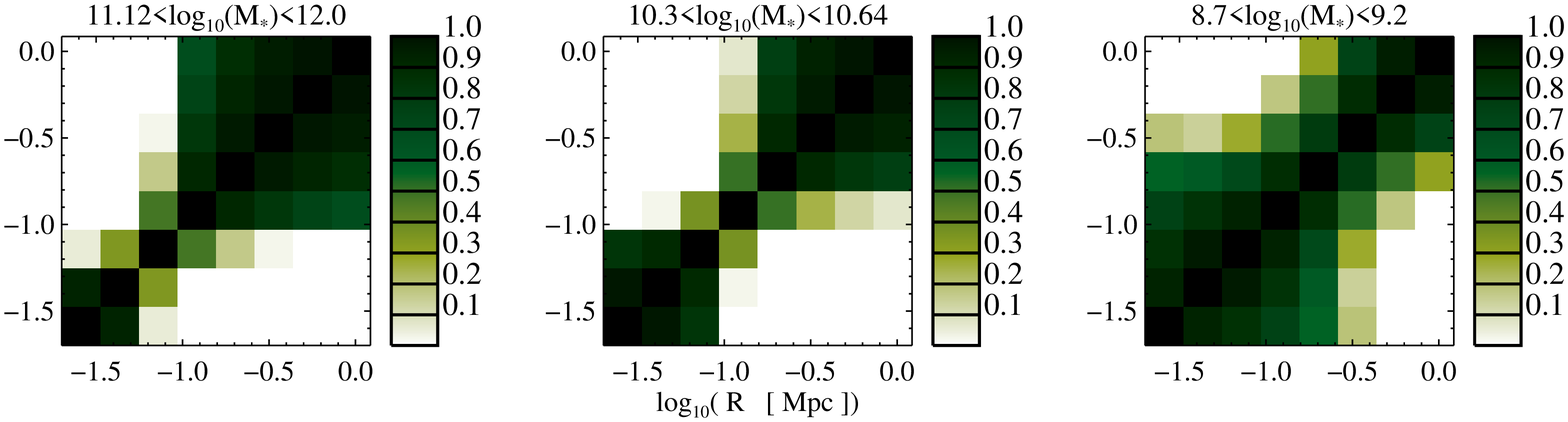}
\caption{Correlation coefficient matrix for g-g lensing for
  $0.22<z<0.48$ and for several bins in stellar mass.}
\label{z1_lensing_covar_matrix}
\end{figure*}

Figure \ref{gg_sample_variance} illustrates the effect of sample
variance on g-g lensing signals for various stellar mass bins and for
$0.22<z<0.48$. Figure \ref{z1_lensing_covar_matrix} shows the
associated correlation coefficient matrices. The key point to note
here is that the sample variance for g-g lensing is dominated by the
one-halo satellite term on scales of about 100 kpc to 1 Mpc. The
impact of this term becomes more apparent in galaxy samples with lower
stellar masses as the contribution from the one-halo central term
decreases. The fact that the one-halo satellite term has a large
sample variance compared to the one-halo central term can be
understood as follows. Consider a sample of galaxies in a given
stellar mass bin. The galaxies that are satellites in this sample will
tend to live in more massive halos than the galaxies that are centrals
(this can be seen in Figure \ref{hod_bins_illustration} for
example). Since more massive halos are more rare than less massive
halos at fixed survey volume, this explains the large one-halo
satellite sample variance.

\begin{figure*}[htb]
\epsscale{1.0}
\epsscale{0.8} 
\plotone{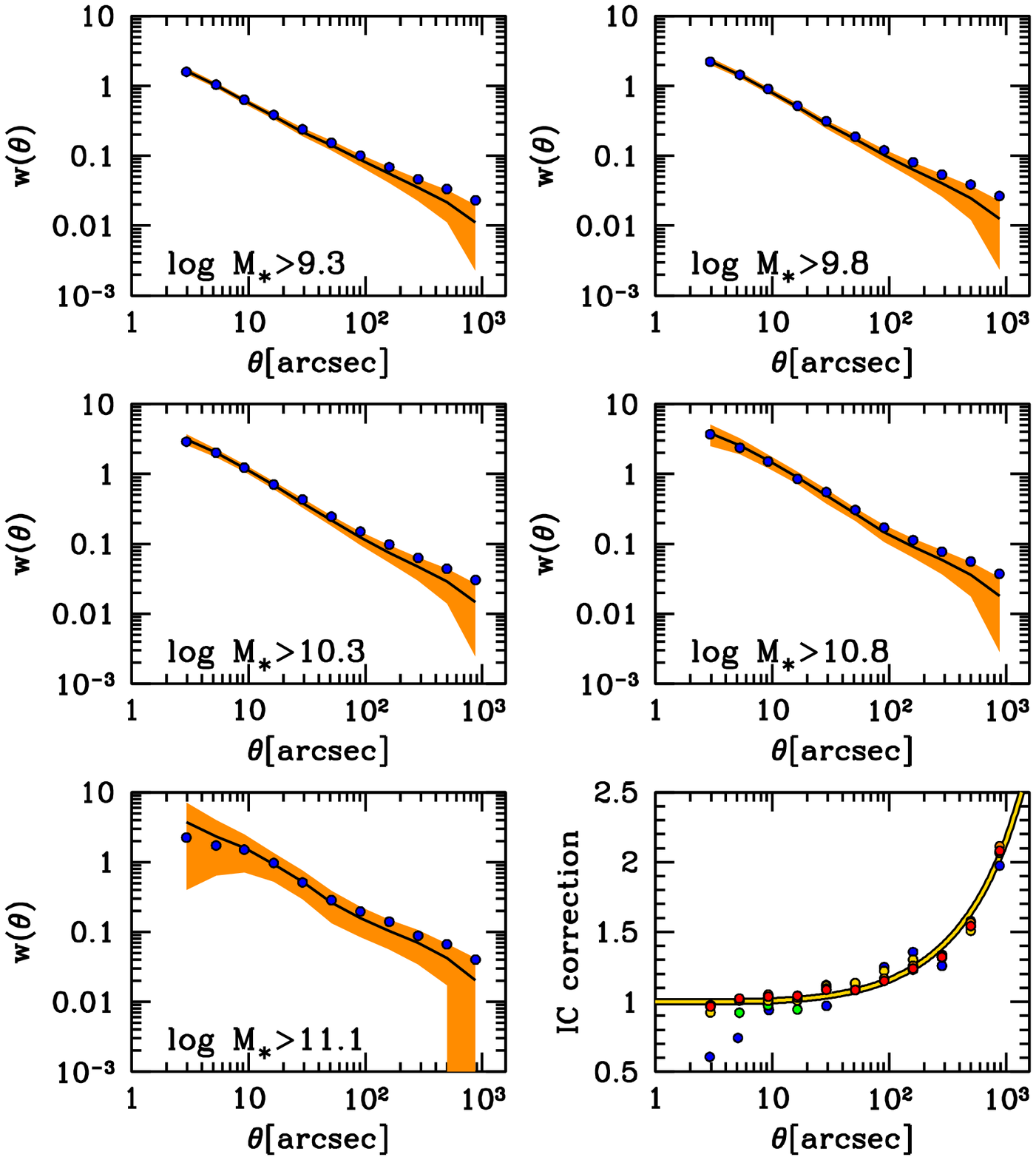}
\caption{The mean and dispersion of the angular clustering of galaxies
  as a function of stellar mass threshold in our mock COSMOS
  simulations. The results shown here are for $0.48<z<0.74$. The
  filled circles in each panel show $w(\theta)$ for a single mock with
  $\gtrsim 10\times$ the area as COSMOS itself. The difference between
  the large-area mock and the mean of the COSMOS mocks is due to the
  integral constraint. The bottom right panel shows the ratio of
  $w(\theta)$ for the large-area mock divided by the mean of the
  COSMOS mocks. The solid curve is a fitting function to account for
  the IC. For the $0.22<z<0.48$, the effect of the integral constraint
  is stronger, while for $0.74<z<1.0$ there is sufficient volume such
  that the IC is unity on all scales measured.}
\label{ic_correction}
\end{figure*}

Finally, we also use mock catalogs to estimate the effects of the
integral constraint (IC) \citep[][]{Groth:1977} on clustering
measurements for a small area survey. Due to spatial fluctuations in
the number density of galaxies, the mean correlation function measured
from an ensemble of samples will be smaller than the correlation
function measured from a single contiguous sample of the same volume
as the sum of the ensemble sample. This attenuation of $w(\theta)$
becomes relevant on angular scales significant with respect to the
sample size. For large surveys like the SDSS, the IC is not an issue
on scales of interest. For a pencil-beam survey like COSMOS, however,
the IC must be taken into account when modeling the clustering. We
estimate the IC correction to our $w(\theta)$ measurements through the
use of the mock galaxy distributions described previously. The results
are shown in Figure \ref{ic_correction}. For COSMOS, our fitting
functions for the IC correction are:

\begin{equation}
  f_{IC} = \exp(\log_{10}(\theta)/3.4)^{2.5},
\end{equation}  
 
\begin{equation}
f_{IC} = \exp(\log_{10}(\theta)/3.2)^{4.1},
\end{equation}

\noindent for $0.22<z<0.48$ and $0.48<z<0.74$ respectively and where
$\theta$ is expressed in arc-seconds. For $0.74<z<1$ there is
sufficient volume such that $f_{IC}=1$. We note that these fitting
functions are only valid for $\theta<10^3$ arcseconds.

\section{Summary and conclusions}\label{conclusions}

The goal of this paper is to develop the theoretical framework
necessary to combine measurements of galaxy-galaxy lensing, galaxy
clustering, and the galaxy stellar mass function, into a single and
more robust probe of the galaxy-dark matter connection. We have
achieved this goal by introducing several key modifications to the
standard HOD framework. To begin with, we have modified the standard
HOD model so as to fit all three probes simultaneously and
independantly of the selected binning scheme. Next, since we are
interested in the galaxy-dark matter connection, we have also modified
the HOD model so as to specifically include the stellar-to-halo mass
relation (SHMR). In a companion paper, Leauthaud et al. 2011b, we
demonstrate that the model presented here provides an excellent fit to
galaxy-galaxy lensing, galaxy clustering, and stellar mass functions
measured in the COSMOS survey from $z=0.2$ to $z=1.0$.

Nonetheless, while the promise of combined dark matter probes in
studying galaxy formation, gravity and cosmology is clear, we must
ensure that our parametric description of the SHMR is sophisticated
enough to capture its possible behavior. There are a number of
questions that remain to be answered in order to achieve this
goal. For example, is $P(M_*|M_h)$ well described by a log-normal
distribution and is the scatter in $P(M_*|M_h)$ constant or does it
vary with halo mass? Do the parameters that describe $P(M_*|M_h)$ vary
with redshift and galaxy type? Can we marginalize over uncertainties
related to the shapes and concentrations of dark matter halos? What
exactly do we learn from various probe combinations? The challenges
are steep but with increasing large data-sets such as the Dark Energy
Survey, the Large Synoptic Survey Telescope, the HyperSuprime Cam
survey, and EUCLID, refined and sophisticated models can be built and
constrained by the data. Although the model presented in this paper is
sophisticated enough to describe COSMOS data, it is clear that further
refinements will be necessary given the statistical precision of
up-coming surveys. Improving models such as the one presented in this
paper by using insights provided, for example, by semi-analytic models
of galaxy formation and dark matter N-body simulations, is clearly a
worthy pursuit.


\acknowledgments
\noindent {\bf Acknowledgments}

We thank Jaiyul Yoo for help with the Appendix and Kevin Bundy for
useful discussions and for reading the manuscript. AL acknowledges
support from the Chamberlain Fellowship at LBNL and from the Berkeley
Center for Cosmological Physics. This research received partial
support from the U.S. Department of Energy under contract number
DE-AC02-76SF00515.  RHW and PSB received additional support from NASA
Program HST-AR-12159.A, provided through a grant from the Space
Telescope Science Institute, which is operated by the Association of
Universities for Research in Astronomy, Incorporated, under NASA
contract NAS5-26555.  MTB and RHW also thank their collaborators on
the LasDamas project for critical input on the Consuelo simulation,
which was performed on the Orange cluster at SLAC.
 
 
 


 

\appendix

\section{Further details on the origin of Equation 34}

We consider it useful to provide some more details on the origin of
Equation \ref{eq:1h} and in particular, on the link between the NFW
profile and $F'_{c}$ and $F'_{s}$. This might be useful for those who
are not familiar with the notations of galaxy clustering studies. To
begin with, consider a central galaxy that is associated with a NFW
dark matter halo at redshift $z_L$ and with halo mass $M$. The NFW
profile, $\rho_{\textsc{nfw}}$, is given by:

\begin{equation}
  \frac{\rho_{\textsc{nfw}}(r, z_L)}{\rho_{\rm crit}} = \frac{\delta}{(r/R_s)\left(1+r/R_s\right)^2},
\end{equation}

\noindent where $\delta$ is a characteristic (dimensionless) density
and $R_s$ is the NFW scale radius. The relation between $\delta$ and
the NFW concentration parameter $c$ is:

\begin{equation}
  \delta=\frac{\Delta}{3}\times\frac{c^3}{\ln(1+c)-c/(1+c)},
\label{delta_nfw}
\end{equation}

\noindent where $\Delta$ is a chosen over-density (for example,
$\Delta$ is often set to 200). The NFW radius is noted
$R_h$ and is equal to $R_h=c\times R_s$. The
projected surface mass density of this lens, $\Sigma$, is computed by
taking the integral of $\rho_{\textsc{nfw}}$ over the line-of-sight:

\begin{equation}
  \Sigma_{\rm co}(r_{\rm p,co},z_L|M) = \int
  \rho_{\textsc{nfw}}\left(\sqrt{r_{\rm p,co}^2+r_{||,\rm co}^2},z_L\right){\rm
    d}r_{\rm ||,co}. 
\label{nfw_projection}
\end{equation}

Analytical expressions for the projection of $\rho_{\textsc{nfw}}$ to
$\Sigma$ can be found in \citet[][]{Wright:2000} for example.

Instead of a single galaxy, now consider an ensemble of central
galaxies characterized by the central occupation function $\ncen$. We
note $\overline{n}_c$ such that:

\begin{equation}
  \overline{n}_{c} \equiv \int \ncen \frac{{\rm d}n}{{\rm d}M} {\rm d}M.
\end{equation}

The probability that a galaxy in this selection lives in a halo of mass $M$ is:

\begin{equation}
  P(M) =  \ncen \frac{1}{\overline{n}_{c}} \frac{{\rm d}n}{{\rm d}M}.
  {\rm d}M.
\end{equation}

The average surface mass density of the galaxy ensemble is:

\begin{eqnarray}  
  \Sigma_{\rm co}(r_{\rm p,co},z_L) &=& \int \int  \Sigma_{\rm
    co}(r_{\rm p,co},z_L|M) P(M) {\rm d}M{\rm d}r_{\rm ||,co} \nonumber\\
  &=& \int \int   \rho_{\textsc{nfw}}\left(\sqrt{r_{\rm
        p,co}^2+r_{||,\rm co}^2},z_L\right) \ncen
  \frac{1}{\overline{n}_{c}} \frac{{\rm d}n}{{\rm d}M} {\rm d}M{\rm d}r_{\rm ||,co}.
\label{app_eq1}
\end{eqnarray}

Let us now define $F'_{c}$ such that:

\begin{equation}
\rho_{\textsc{nfw}}(r, z_L) = \frac{1}{4\pi r^2}\times M \times \frac{1}{2R_{h}}\times F'_{c}\left(\frac{r}{2R_{h}}\right).
\label{app_eq2}
\end{equation}

Combining Equation \ref{app_eq1} and Equation \ref{app_eq2} we obtain:

\begin{equation}
  \Sigma_{\rm co}(r_{\rm p,co},z_L) = \frac{1}{4\pi r^2
    \overline{n}_c}\int \int \frac{{\rm d}n}{{\rm
      d}M}\frac{M}{\overline{\rho}}\frac{1}{2R_{h}}n_{\rm tot}F'_{c}\left(\frac{\sqrt{r_{\rm p,co}^2+r_{||,\rm
          co}^2}}{2R_{h}}\right) {\rm d}M{\rm d}r_{\rm ||,co}.
\end{equation}

Finally, Equation \ref{eq:1h} is obtained by considering a galaxy
sample that contains both central and satellite galaxies. In this case
$\overline{n}_g$ is defined as:

\begin{equation}
  \overline{n}_{g} \equiv \int \left(\ncen+\nsat\right) \frac{{\rm d}n}{{\rm d}M} {\rm d}M.
\end{equation}

\noindent $F'_{s}$ is defined in a similar fashion to Equation
\ref{app_eq2} but $\rho_{\textsc{nfw}}$ is replaced with the
convolution of the NFW profile with itself. Analytic expressions for
the convolution  of the truncated NFW profile with itself can be found
in the Appendix of \citet[][]{Sheth:2001a}.

 
 
 

\bibliographystyle{apj}


\end{document}